\documentclass[traditabstract]{aa} 
\usepackage{graphicx}
\usepackage{txfonts}
\usepackage{textcomp}

\usepackage{color}
\usepackage{natbib}
\bibpunct{(}{)}{;}{a}{}{,} 
\usepackage{hyperref}
\usepackage{longtable}
\usepackage{txfonts}

\usepackage{rotating}
\usepackage{amssymb}
\usepackage{url}
\begin{document}
  \title{C and N abundances of MS and SGB stars in NGC~1851.\thanks{Based on observations taken with the 6.5 meter Magellan 
Telescope at Las Campanas Observatory, Chile and with the Very Large 
Telescope at ESO La Silla Paranal Observatory, Chile, under programme ID 
68.D-0510.}}


   \author{C. Lardo\inst{1}, A. P. Milone\inst{2,3},
	  A. F. Marino\inst{4}, A. Mucciarelli\inst{1}, E. Pancino\inst{5}, M. Zoccali\inst{5,6}, M.Rejkuba\inst{7}, R. Carrera\inst{2,3} \and O. Gonzalez\inst{7}.  
          }
  
  \authorrunning{Lardo et al.}

   \institute{Department of Astronomy, University of Bologna, Via Ranzani 1,
 40127 Bologna, Italy;~\email{carmela.lardo2@unibo.it}\\
       	\and Istituto de Astrof\`\i sica de Canarias, via Lactea a/n, 38200, La Laguna, Tenerife, Spain\\
	\and Department of Astrophysics, University of La Laguna, E-38200 La Laguna, Tenerife, Canary Islands, Spain\\
	\and Max-Planck-Institut f\"{u}r Astrophysik, Postfach 1317, D85748 Garching b. M\"{u}nchen, Germany\\
	\and INAF-Osservatorio Astronomico di Bologna, Via Ranzani 1, 40127 Bologna, Italy\\
        \and Pontificia Universidad Cat\'{o}lica de Chile, Departamento da Astronom\'{\i}a y  Astrof\'{\i}sica, Casilla 306, Santiago22, Chile\\
	\and ESO, Karl-Schwarzschild-Strasse 2, D-85748 Garching b. M\"unchen, Germany\\
        }

   \date{Received/Accepted}

 \abstract
 {

We present the first chemical analysis of stars on the double subgiant branch (SGB) of the globular cluster NGC~1851.
We obtained 48 Magellan IMACS spectra of subgiants and fainter stars covering the spectral region between 3650-6750\AA, 
to derive C and N abundances from the spectral features at 4300\AA~($G$-band) and at $\sim$ 3883\AA~(CN).
We added to our sample $\sim$ 45 unvevolved stars previously observed with FORS2 at the VLT. These two datasets were homogeneously 
reduced and analyzed.
We derived abundances of C and N for a total of 64 stars and found considerable star-to-star variations 
in both [C/H] and [N/H] at all luminosities extending to the red giant branch (RGB) base (V$\sim$ 18.9). 
These abundances appear to be strongly anticorrelated, as would be expected from the CN-cycle enrichment, 
but we did not detect any bimodality in the C or N content. 
We used {\it HST} and ground-based photometry to select two groups of faint- and bright-SGB stars from the visual 
and Str\"omgren color-magnitude diagrams.
Significant variations in the carbon and nitrogen abundances are present among stars of each group, 
which indicates that each SGB hosts multiple subgenerations of stars.
Bright- and faint-SGB stars differ in the total C+N 
content, where the fainter SGB have about 2.5 times the C+N content of the brighter ones.
Coupling our results with literature photometric data and abundance determinations from high-resolution studies, 
we identify the fainter SGB with the red-RGB population, which also should be richer on average in Ba and other $s$-process 
elements, as  well as in Na and N, when compared to brighter SGB and the blue-RGB population.}

   \keywords{stars: abundances -- stars: sub giant branch --GCs:
   individual (NGC~1851)-- C-M diagrams}

\maketitle
\section{Introduction}\label{introduzione}

In the last years, both spectroscopic and photometric evidence has shown that globular clusters (GCs)
cannot any longer be considered to be a simple stellar population, because {\em self-enrichment} is a common feature among them.
While the detection of significant scatter in iron and/or $n$-capture elements associated to $s$-process is limited to a
few clusters  \citep[e.g.,][]{hesser82,yong08,marino09}, star-to-star variations in abundances of the light elements
(C, N, O, Na, Mg, Al), have been observed in stars of all evolutionary phases in the majority of 
GCs \citep[e.g.,][]{martell11,pancino10,kraft79,ramirez03,kayser08,carretta09}.

These variations manifest themselves through correlations and anticorrelations, the signature of high-temperature
proton fusion  cycles  that have processed C and O into N, Ne into Na and Mg to Al.
The temperatures required to convert Ne into Na are on the order of T $\sim\times10^{7}$ K and 
these are not reached in $\leq$0.8$M_{\sun}$ GC dwarfs, which, furthermore do not posses a deep convective layer.
The generally accepted explanation is that these stars were {\em born} with the observed CNONa abundance patterns.
Intermediate-mass asymptotic giant branch (AGB) stars, fast rotating massive stars, or massive interactive binaries have 
been proposed as sources of the necessary pollution of the intra cluster medium (ICM)  
before the second-generation stars were formed \citep[see, e.g.][]{dantona07,decressin07,demink09}.

Star-to-star variations in light- and alpha-element abundances, age, and metallicity can determine multimodal or broad sequences in the CMD.
Complex structures along the main sequence (MS), subgiant branch (SGB), red giant branch (RGB) or horizontal branch (HB) within some galactic or extra-galactic GCs (e.g.,~\citealp{pancino00,bedin04,sollima07,piotto07,marino08,milone10,lardo11}; to name a few), 
unambiguously indicate that GCs host two or more generations of stars.

NGC~1851 is one of the most intriguing  clusters with multiple stellar populations. 
It exhibits a double SGB, with the faint component made of $\sim$35\% of stars \citep[][hereafter M08]{milone08}. 
If age is the sole cause, then the SGB split is consistent with two stellar groups with an age difference 
of $\sim$1 Gyr. As an alternative, the two SGBs could be nearly coeval but with a different C+N+O content
\citep{cassisi08,ventura09}. 
The HB is also bimodal, with about $\sim$35\% of HB stars on the blue side of the instability strip. 
Both the HB and the MS morphology leave no room for strong helium variations \citep{salaris08,dantona09}.
Nor is the RGB consistent with a simple stellar population \citep{grundahl99,calamida07}.
\citet{lee09} and \citet{han09} pointed out two {\em distinct} RGB evolutionary
sequences, using Str\"omgren Ca $uvby$ photometry, and proposed that the split might be attributed to differences in 
calcium abundance.

Many spectroscopic studies have been dedicated to RGB stars in NGC 1851.
Almost 30 years ago, \citet{hesser82} noticed three out of eight bright-RGB stars with anomalously strong CN bands 
and with enhanced Sr and Ba lines.
More recently, \citet{yong08} analyzed UVES spectra of eight giants. 
Their analysis revealed that star-to-star abundance variations of O, Na, and Al with a clear anticorrelation of 
O and Na also exist in this cluster, and the amplitude of these variations is comparable with those found 
in clusters of the same metallicity.
More interestingly, they argued for the presence of two different populations in this cluster, characterized 
by significant differences in the  the light 
 $s$-process element Zr and the heavy  $s$-process element La.
 \citet{yong09} and Yong et al.\ (in preparation) found a wide spread in the abundance 
sum C+N+O (while a constant sum of C+N+O was derived by \citealp{villanova10} from the abundance analysis of 15 RGB stars); 
with the CNO-rich stars being also enhanced in Zr and La. 
Yong and collaborators associated the group of CNO-rich $s$-rich stars to the progeny of the faint-SGB and
suggested that intermediate-mass AGB stars might have contributed to the enrichment of the intra cluster medium (ICM) 
before the formation of the second generation of stars.
Interestingly, both the $s$-rich and the $s$-poor groups exhibit their own Na-O anticorrelation, which suggests that 
NGC 1851 has experienced a complex star-formation history. 

\citet{yong08} also suggested the presence of a slight metallicity spread ($\leq$0.1 dex) among NGC~1851 RGB stars. 
This result has been recently confirmed by \citet{carretta10} on the basis of a larger sample of stars.
Following a classification scheme based on Fe and Ba abundance, \citet{carretta10} distinguished between 
a metal-rich, barium-rich (MR) and a metal-poor, barium-poor population (MP).
They associated the MR and the MP components to the bright- and the faint-SGB respectively, 
which is at odds with what was suggested by \citet{yong08}.

While the above listed spectroscopic studies targeted evolved stars in NGC 1851, before this work, only \citet{pancino10}
analyzed a sample of unevolved stars in this clusters to provide index measurements, and we present for the first time 
an abundance analysis of MS, TO and SGB stars.
In particular, we observed stars located on the faint- and bright-SGB and derived for them C and N abundances; 
aiming for the first time to provide insights on the chemical signature differences between the two discrete sequences on the SGB.
The paper is organized as follows:
in Sect.~\ref{observations}  we discuss the observations and data reductions; in Sect.~\ref{definizione_indici} we define 
the index passbands and present our results from index measurements in Sect.~\ref{analisi}. Section~\ref{moog} contains 
a description of model atmosphere parameters and abundance derivations; Sect.~\ref{abbond} presents the abundance results;
Sect.~\ref{cmds} focuses on  C and N abundances of stars strictly located on the faint-SGB or bright-SGB. 
Finally, we summarize our findings and draw the  conlusions in Sect.~\ref{conclusioni}.\\

\section{Observations and data reduction}\label{observations}

\subsection{Source catalogs and sample selection}

\begin{figure}
\resizebox{\hsize}{!}{\includegraphics{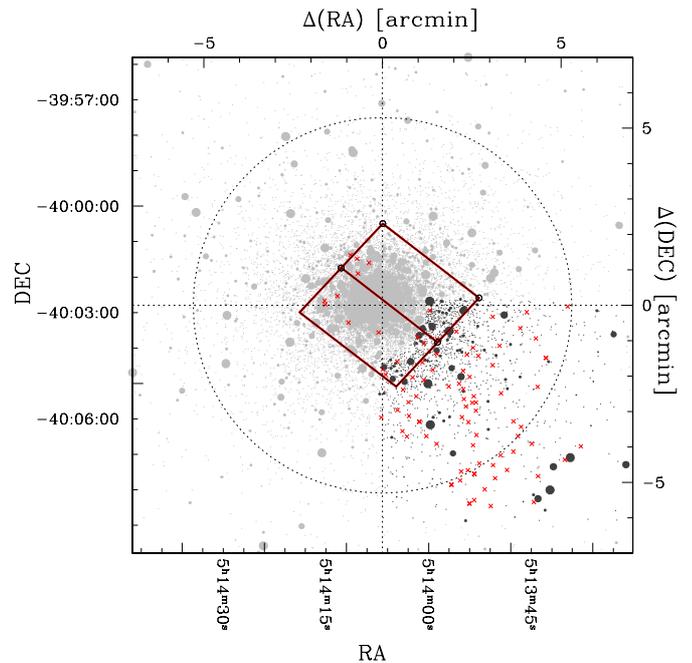}}
\caption{Area covered by the catalogs from which the spectroscopic selection was made: 
the FORS2 in the outer part (shown in black) and the ACS/HST field (continuous red line) in the inner region.
The red crosses represent the IMACS+FORS2 sample. We also indicate the half tidal radius in this figure (dotted circle).}
        \label{area}
   \end{figure}

We selected our targets from literature photometry: FORS2 $V$ and $I$ photometry 
presented by \citet{zoccali09}, in the southwest
quadrant of the cluster, as well as $F606W$ and $F814W$ HST/ACS photometry from the GGC treasury program GO-10775 for the inner part of the cluster (M08).
The area covered by the catalogs and the selected spectroscopic targets is shown in Fig.~\ref{area}.
We transformed the coordinates using 2MASS as a reference astrometric catalog, so the final catalog is on 
the same relative astrometric system.
Spectroscopic targets were selected as the most isolated stars located 
around the turn-off and the SGB, reaching the RGB base.
The resulting photometry was calibrated
using stars in common with the $V,I$ \citet{bellazzini01} catalog, covering an 8'$\times$8' field centered on the cluster.
We also used publicly available Str\"omgren $u,v,b,y$ photometry\footnote{\footnotesize{\tt http://www.mporzio.astro.it/spress/stroemgren.php}.}  of NGC~1851 from \citet{grundahl99} and \citet{calamida07}.
We refer the reader to that paper for details about observations and data reduction of Str\"oemgren photometry.

\subsection{Observations and spectroscopic reductions}
We acquired low-resolution (R$ \simeq$1123 and R$\simeq$1246 at 3880 and 4305~\AA, respectively) 
spectra of turnoff and SGB stars in the globular cluster NGC 1851 with the IMACS multi-object 
spectrograph at the Magellan 1 (Baade) telescope at the Las Campanas Observatory in Chile. 
The adopted instrumental setup with the grating GRAT 600-I covers the nominal spectral range between 
3650-6750~\AA with a dispersion of 0.38\AA~/pix. 
This spectral range includes CN (3880~\AA) and CH (4305~\AA) 
molecular bands used to derive nitrogen and carbon abundances. 
However, the actual spectral coverage depends on the location of 
the slit on the mask with respect to the spectral dispersion. 
Because we observed faint stars, the total integration time was 
long, requiring ten exposures of 1800 sec each. Therefore all 
observations were made with a single-mask setup with 48 slits. 
We were able to extract spectra for 46 targets from our initial target list. 

To these 46 spectra observed with Magellan, we added 47 other 
MS and SGB spectra from \citet{pancino10} observed with the FORS2 multi-object spectrograph on 
the ESO VLT at the Paranal Observatory in Chile. We refer the reader to that paper for details of the FORS2 observations.
We reduced our data following the procedure described in \citet{pancino10}.
For the data pre-reduction, we used the standard procedure for overscan correction
and bias-subtraction with the routines available in the \textit{noao.imred.ccdred} package in
IRAF\footnote{IRAF is distributed by the National Optical Astronomy Observatory,
which is operated by the Association of Universities for Research in Astronomy, Inc., under cooperative agreement with the National
Science Foundation.}. Cosmic rays were removed with the IRAF Laplacian edge-detection routine \citep{vanDokkum01}.
The frames were then flat-fielded and reduced to one dimension spectra with the task \textit{apall}.
Once  we obtained ten wavelength-calibrated, one-dimensional spectra for each star, we co-added them on a star-by-star 
basis to reach a relevant S/N (typically between 20-30 per pixel at 3880\AA) even in the bluer part of the spectrum
\footnote{Before co-adding we checked that the shifts between spectra from different exposures were 
negligible compared to our wavelength calibration uncertainty (30 km s$^{-1}$) .}.
As a final step, we examined each spectrum and rejected those spectra with bad quality, following the recipes
outlined in \citet{pancino10}.
We defined several criteria for this rejection:
\begin{enumerate}
 \item S/N ratio $<$ 10 (per pixel) in the CN 3880\AA~band;
 \item clear defects (like spikes or holes) from an individual inspection of the spectrum on the band or continuum windows;
 \item discrepant Ca(H+K) and $H_{\beta}$ index measurements.
\end{enumerate}
Forty-three stars survived our criteria of selection.  Moreover, 20 stars had spectral or continuum passbands falling in the gap between the CCDs because of the location of the slit on the mask with respect to the dispersion direction. We were not able to measure CH and CN band strengths for these stars, but in some cases we determined N and C abundances (Sect.~\ref{moog}). 
Therefore we decided to retain these spectra and consider them in the subsequent analysis.

\begin{figure}
\resizebox{\hsize}{!}{\includegraphics {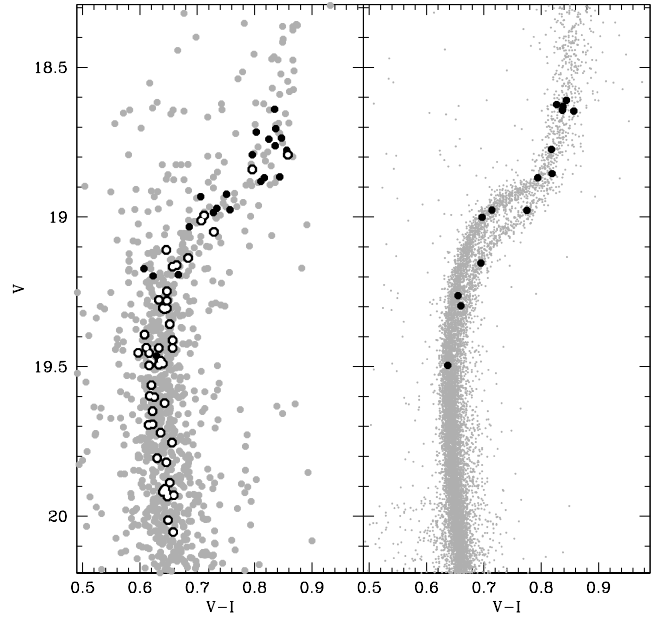}}
\caption{Color-magnitude diagrams for NGC~1851. Gray dots show 
$V-I$ photometry for the outer field 
({\em left panel}) and inner field ({\em right panel}) from FORS2 and ACS, respectively.
White dots mark spectroscopic targets presented in \citet{pancino10}, while black dots 
show our newly observed stars.}

        \label{cmd}
   \end{figure}

\subsection{Membership and quality control}
NGC 1851 \citep[$l,b=244.51,-35.08$;][2010 edition]{harris96} is projected at a far distance from the Galactic 
plane and consequently the contamination from field stars is almost negligible.
In addition, the average radial velocity of cluster stars significantly differs with 
respect to the field \citep[320.5 km/s;][2010 edition]{harris96}, hence non cluster members 
can be easily identified from their radial velocities.
For the stars in the ACS/WFC field of view, member stars were furthermore selected on the 
basis of their proper motions (see M08 for details).

For all the spectra, radial velocities were measured with the IRAF task {\em fxcor},
which performs the Fourier cross-correlation between 
the object spectrum and a template spectrum (the latter with known radial velocity). 
As a template we chose the spectrum of a star with the highest S/N: its radial velocity was computed using the laboratory positions of several strong lines 
(e. g. $H_{\beta}$, $H_{\alpha}$, $H_{\gamma}$ and $CaHK$ among others) with the IRAF task {\em rvidlines}.
To derive a robust determination for the radial velocities, we performed for a given star 
the cross-correlation against the template in four different spectral regions that
span the entire spectral coverage, from the bluest part out to the reddest part of the spectrum.
Then, the four values were averaged together, obtaining typically internal errors of
$\sim$ 25-30 $km s^{-1}$.
We obtained an average $V_{rad}$ of 317 $km s^{-1}$ with a dispersion of 11 $km s^{-1}$, which fully agrees with the 
previous determination by \citet{yong08}, \citet{villanova10}, and \citet{carretta10}. 

\subsection{Additional literature data}
\label{stelle_elena}
To increase our observed sample, we added 47 MS and SGB stars observed with FORS2 that were 
presented by \citet{pancino10}.
The typical resolution of these spectra is R=$\lambda / \delta \lambda$$\simeq$800 and the S/N ratio (per pixel) in the 
CN band region is S/N(CN)$\simeq$30.
More details concerning the observations and data reduction can be found in \citet{pancino10} and references therein.
We reduced our data following the same procedure as adopted by these authors.
Moreover, their definition of indices is exactly the same as ours (Sect.~\ref{definizione_indici}).
While \citet{pancino10} did not need to normalize their spectra, this was necessary in our case, so we fitted a polynomial to the pseudo continuum to normalize both IMACS and FORS2 spectra as described in Sect.~\ref{definizione_indici}.
We adopted the same procedure in determining the C and N abundance for both the newly obtained and previously studied spectra. 
 
 \section{Index definition and measurements}\label{definizione_indici}
Spectral indices are defined as a window centered on the molecular band of interest
and one or two windows around it to define the continuum level.
For each spectrum, S3839 and CH4300 indices sensitive to the absorption by the 3883\AA~CN band and the
4300\AA~CH band were measured. 
Several spectral index definitions exist in literature, generally optimized to quantify the CN content of the atmospheres
of red giant stars.
In our case, to be consistent with the previous work of \citet{pancino10}, we decided to adopt the indices as 
defined in \citet{harbeck03}:
$$S3839(CN) = -2.5 \log \frac{F_{3861-3884}}{F_{3894-3910}}$$
$$S4142(CN) = -2.5 \log \frac{F_{4120-4216}}{0.5F_{4055-4080} + 0.5F_{4240-4280}}$$
$$CH4300 = -2.5 \log \frac{F_{4285-4315}}{0.5F_{4240-4280} + 0.5F_{4390-4460}}.$$
In particular, the S3839(CN) index we used differs from that of \citet{norris81} or \citet{norris79}, 
and it accounts for stronger hydrogen lines in the region of CN feature for stars cooler than red giants.
As an additional check, we defined and measured two different CN band indices in the wavelength region covered by our spectra: 
the S3839 for the CN band around 3880~\AA~and the S4142 for the one around 4200~\AA\footnote{
As discussed by different previous authors, the S3839 index is found to be by far much more sensitive to 
CN variations with respect to S3839. For the S4142 index, the spread is generally of the size of (or slightly wider) 
than the median error bar on the index measurements. Therefore, we decided to rely only on the S3839 index measurements throughout.}.
We obtained index measurement uncertainties with the expression derived by \citet{Vollmann06}, 
assuming pure photon noise statistics in the flux measurements.
In addition, we measured the two indices centered around the calcium H and K lines and the $H_{\beta}$ line as 
defined again in \citet{pancino10} to reject outliers from our sample.

The  final reduced spectra generally show a strong decline of the signal toward bluer wavelengths.
This is largely expected and may be due to the different instrumental efficiency, higher absorption of the Earth's 
atmosphere in the blue and stellar flux wavelength dependency \citep[see][for a complete discussion]{cohen02}.
The CH band at ~4300~\AA\, is not affected by the change in spectral slope from atmosphere or instrumental effects
thanks to two continuum bandpasses.
On the other hand, we had to rely only on a
single continuum bandpass in the red part of the spectral feature for the 3883\AA~CN band.
Following \citet{cohen02,cohen05}, we decided to normalize the stellar continuum in the spectrum of each star, then found the 
absorption within the CN bandpass. Moreover, by fitting the continuum, we were able to directly  
compare the indices measured 
in this section and the abundances derived from spectral synthesis in Sect.\ref{moog}.
We fitted a third-order polynomial masking out the region of the CN band \citep[see][]{cohen05}. 
The polynomial fitting used a 6$\sigma$ high and 3$\sigma$ low clipping, running over a five pixel average.
Then we computed S3839 indices from these continuum-normalized spectra and used the average (0.126$\pm$0.04 and 0.05$\pm$0.01 for IMACS and FORS2 spectra respectively) to set a zero point offset and thus delete the instrumental signature present in the raw S3839 indices.

The measured indices are listed in Tab.~\ref{indici_tab} and plotted in 
Fig.~\ref{ridge}. 

\addtocounter{table}{1}

\begin{figure}
\resizebox{\hsize}{!}{\includegraphics{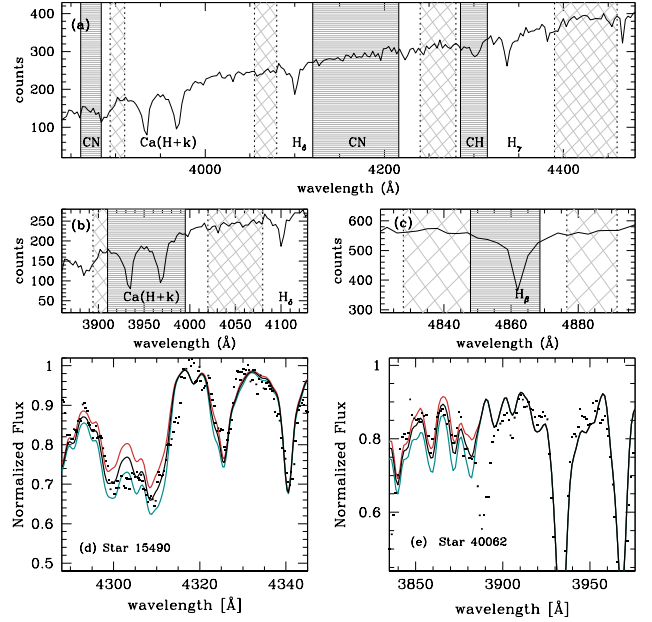}}
\caption{{\em Panel a}: example of the windows from which we measured the CN and CH indices 
(dark gray hatched regions) together with their respective
continuum windows (light gray hatched regions).
{\em Panels b} and {\em c} show the windows adopted for the H and K calcium index and the 
$H_{\beta}$ index. 
The non-normalized superimposed spectrum (star 41213, S/N$\sim$35 in the S3839 region) was smoothed for clarity.
{\em Panel d}: Observed (small black dots) and synthetic (line) spectra around CH band for the star 15490. 
The black best fits, 
while the red and green lines are the syntheses computed with C abundance altered by $\pm$ 0.10 dex from 
the best value.
{\em Panel e}: The same as in {\em panel d} but for the CN feature for the star 40062. The synthetic spectra 
show the best fit (thick black line) and the syntheses computed with N abundance altered by $\pm$ 0.20 (thick red and green lines).} 
        \label{spectrum}
   \end{figure}

\begin{figure}
\resizebox{\hsize}{!}{\includegraphics{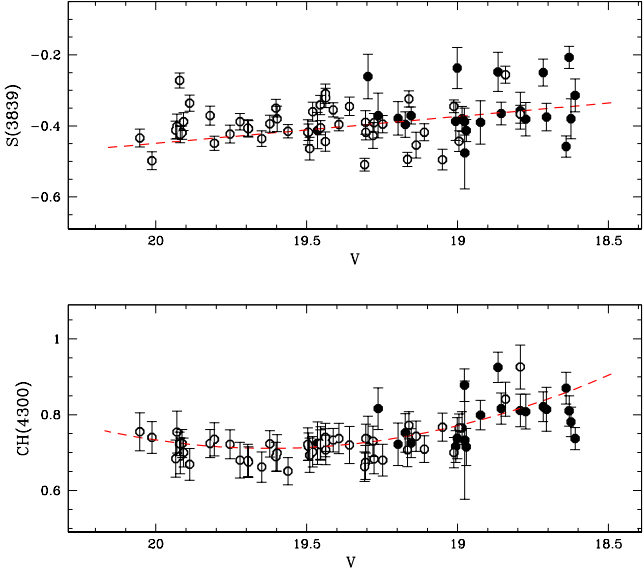}}
\caption{Measured S3839 and CH4300 indices for the program stars as a function of $V$. White dots are stars from \citet{pancino10}, while black dots are from the IMACS sample. Dashed red lines in both panels are the {\em median ridge lines} used to eliminate temperature and gravity dependencies.}
        \label{ridge}
   \end{figure}

\subsection{Dependency on temperature and gravity}\label{temp_gravita}
CN and CH bands are stronger at a fixed overall abundance for stars with lower temperature and gravity.
In particular, the formation efficiency of the CN molecule strongly depends on the temperature, 
therefore we expect the indices depend on the color of the single stars.
These dependencies are usually eliminated 
\citep[see][for example]{harbeck03,kayser08} by fitting the lower envelope of the distribution in the index-magnitude plane 
(or index-color plane). For our  sample, we used the {\em median ridge line}, 
shown as dashed red lines in Fig.~\ref{ridge}, to eliminate these dependencies \citep[see][]{pancino10}.
These rectified CN and CH indices are in the following indicated as $\delta$S3839 and $\delta$CH4300, respectively, and
we refer to these new indices throughout\footnote{We obtained a rough estimate of the uncertainty 
in the placement of these median ridge lines by using the first interquartile of the rectified indices divided by the 
square root of the total points.
The uncertainties (typically $\sim$ 0.01 for the CN index and $\sim$ 0.005 for the CH index) 
are largely negligible for the applications of this work.}.
For the stars in common between this work and \citet{pancino10}, the mean difference in the $\delta$S3839 and $\delta$CH4300 indices,
derived by subtracting Pancino et al.'s values from ours, are 0.00 $\pm$ 0.02 mag and --0.01$\pm$0.01 mag, respectively.
Because we adopted the same reduction procedures as in \citet{pancino10}, we can only ascribe this small difference 
to the continuum normalization we performed (see Sect.~\ref{definizione_indici}).

\subsection {CN and CH distribution}\label{analisi}
Variations of several light elements and anticorrelations between strengths of the CN and CH bands 
were detected for very many clusters.
Because molecular abundance is controlled by the abundance of the minority species, the corrected
CH index is a {\em proxy} for carbon abundance, while $\delta$S3839 traces the nitrogen abundance.
The visual inspection of the top panel of Fig.~\ref{ridge} reveals significant scatter in the CN index
over the magnitude range with $V \gtrsim$ 19.5 with some hints of bimodality toward the brightest tail of the distribution.  
The range of CN becomes less evident at fainter luminosities as a consequence of the increasing temperature.
In the bottom panel of the same figure, we plot the CH4300 and S3839 versus the 
stellar V-band magnitudes. Here the variations among the measured index are very small and within the uncertainties throughout. 

\begin{figure}
\resizebox{\hsize}{!}{\includegraphics{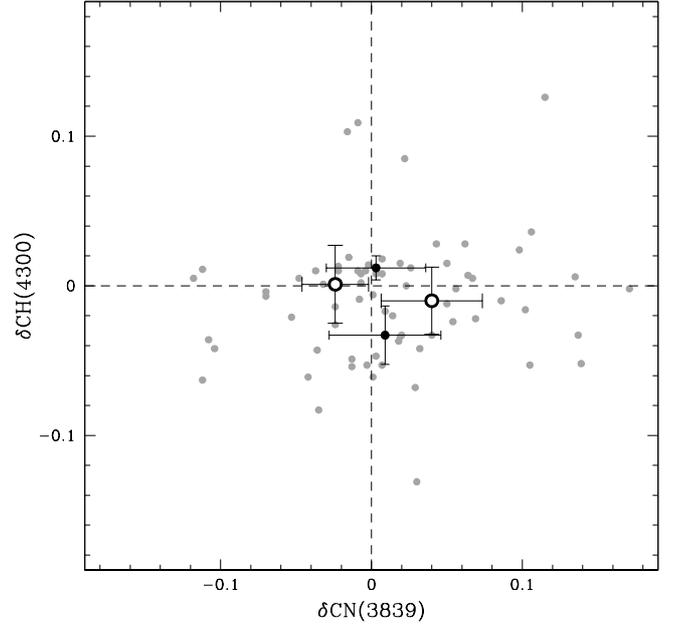}}
\caption{Plot for the distribution of CN and CH band strengths.
Gray dots show measurements for stars. CH-weak and CH-strong stars are separated by the horizontal dashed line and their 
centroids with error bars (drawn at 1$\sigma$) are marked as black dots and large empty dots, respectively.}
        \label{isto}
\end{figure}

Figure~\ref{isto} shows the rectified index  $\delta$S3839 as a function of $\delta$CH4300 for all stars. 
We found no evidence for a significant CH-CN anticorrelation, similarly to what was found by \citet{pancino10}.

\section{Spectral synthesis and abundance derivations}\label{moog}
Indices are a fast tool to characterize chemical anomalies,
but we can also rely on spectral synthesis to fully characterize our target stars.
This becomes necessary when indices do not offer conclusive answers, 
as we saw in the previous sections.

\subsection{Atmospheric parameters}\label{parametri}
We derived estimates of the atmospheric parameters from the calibrated ACS and FORS2 photometry presented
in Sect.~\ref{observations}.
Dereddened $(V-I)_{0}$ colors were obtained adopting $E(B-V)$ = 0.02 \citep[][2010 edition]{harris96}.
We obtained effective temperatures and bolometric corrections (hereafter $T_{eff}$ and $BC_{V}$) 
with the \citet{alonso96,alonso99,alonso01} color-temperature relations, adopting [Fe/H]= -1.22 from \citet{yong09},
and taking into account the uncertainties in the magnitudes and reddening estimates.
\citet{alonso96,alonso99} adopted Johnson's system as a reference, therefore we converted $(V-I_{C})$ into $(V-I_{J})$
after dereddening using the prescriptions by \citet{bessell79} to feed the \citet{alonso96,alonso99,alonso01} calibration.
Gravities were then obtained by means of the fundamental relations

$$ \log \frac{g}{g_{\sun}} = \log \frac{M}{M_{\sun}} +2\log \frac{R_{\sun}}{R},$$
$$ 0.4~(M_{bol}-M_{bol,\sun}) = -4 \log \frac{T_{eff}}{T_{eff,\sun}} +2\log \frac{R_{\sun}}{R},$$

where we assumed the solar values reported in \citet{andersen99}: $\log{g_{\sun}}$ = 4.437,
$T_{eff,\sun}$ = 5770K and $M_{bol,\sun}$ = 4.75. 
For all our stars, we assumed a typical mass of 0.8 $M_{\sun}$ \citep{berg01} and 
a distance modulus of $(m-M)_{V}$ = 15.47 \citep[][2010 edition]{harris96}.

\begin{figure}
\resizebox{\hsize}{!}{\includegraphics{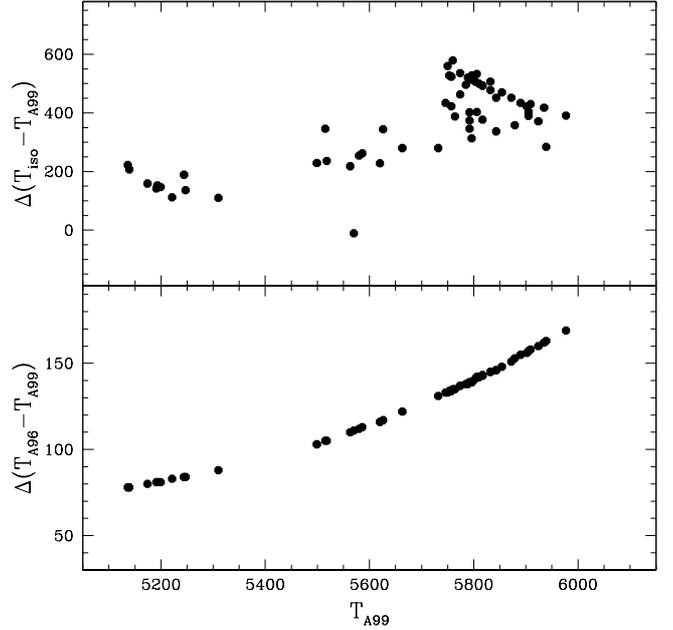}}
\caption{ {\em Top panel}: Differences in the temperature estimates by means of isochrone projection procedure and using the \citet{alonso99} relation 
($\Delta(T_{iso}-T_{A99}$)) as a function of the \citet{alonso99} temperature ($T_{A99}$) for all our target stars. {\em Bottom panel}: The same,
but for the  \citet{alonso96} calibration ($T_{A96}$).}
      \label{temp}
\end{figure}

Finally, we obtained the microturbulent velocities ($v_{t}$) 
from the relation $\log g$ and $v_{t}$, i.e., $v_{t}=1.5-0.03 \log g$ as in  \citet{carretta04}. This method 
leads to an average microturbulent velocity estimate of  $v_{t}=1.0\pm0.1$ km$s^{-1}$, therefore
we assumed $v_{t}=1.0$ km $s^{-1}$ for the entire sample.
An additional check to test the reliability of our atmospheric parameter determination was performed using theoretical isochrones downloaded from the BaSTI\footnote{\url{http://albione.oa-teramo.inaf.it/}} database \citep{pietrinferni06}.
We chose an isochrone of 11 Gyr (12 Gyr for the faint-SGB) with standard $\alpha$-enhanced composition, 
and metallicity Z = 0.002 and we projected our targets on the isochrone to obtain their parameters. 
We present the average difference between the two methods in Fig.~\ref{temp}.
In the top panel we plot the difference in temperature obtained using the \citet{alonso99} calibration and the temperature obtained
by projection of the star on the BaSTI isochrones as a function of the temperature derived by means of the \citet{alonso99} empirical relations.
The scatter for high temperatures is significant, as largely expected, because these empirical calibrations were obtained for giant stars and therefore are valid in 
a precise range of color. On the other hand, the scatter is modest (and in many cases within uncertainties) when comparing temperatures obtained by using the\citet{alonso99} and \citet{alonso96} relations (see bottom panel of Fig.~\ref{temp}), where the latter was derived for low main sequence stars.
However, we preferred to avoid using temperatures derived by isochrone fitting mainly for these reasons: {\em (a)} we cannot assume {\em a priori} that the 
cluster is a single population (with the same [Fe/H] and CNO content, among others) and {\em (b)} the projection on the ($V,V-I$) plane is always uncertain, and a 
rigorous treatment should include (asymmetrical) errors on the $V$ magnitude and $V-I$ color, and finally, {\em (c)} different sets of isochrones (Padova, BaSTI, and DSEP for example) give different results. 

As discussed in the following sections, even if the differences between the two temperatures scales appear non-negligible, the main results of the paper appear totally unchanged if we adopt one or the other temperature scale. This is mainly because the abundances ranking among target stars is left unchanged.
We therefore preferred to rely on  the \citet{alonso99} parameter estimates and discuss the effect of the chosen temperature scale below.

\subsection{Abundances derivation} 

We used the local thermodynamic equilibrium (LTE) program MOOG \citep{sneden73} combined with the ATLAS9 
model atmospheres \citep {kurucz93,kurucz05} to determine carbon and nitrogen abundances.
The atomic and molecular line lists were taken from the latest Kurucz compilation and 
downloaded from F. Castelli's website\footnote{\url{http://wwwuser.oat.ts.astro.it/castelli/linelists.html}}.

Model atmospheres were calculated with the ATLAS9 code 
starting from the grid of models available in F. Castelli's website, 
using the values of $T_{eff}$, $\log g$, and $v_{t}$ determined as explained in the previous section.
The ATLAS9 models employed were computed with the new set of opacity distribution functions
\citep{castelli03} and excluding approximate overshooting in 
calculating the convective flux.
For the CH transitions, the $\log$ g obtained from the Kurucz database were revised downward by
0.3 dex to better reproduce the solar-flux spectrum by \citet{neckel84}
with the C abundance by \citet{caffau11}, as extensively discussed in \citet{mucciarelli11}.

C and N abundances were estimated by spectral synthesis of the 2$\Sigma$--2$\Pi$ band of CH (the G band) 
at $\sim$4310\AA~and the $UV$ CN band at 3883\AA~(including a number of CN features in the wavelength
range of 3876-3890\AA), respectively. 
Lower panels of Fig.~\ref{spectrum} illustrate the fit of synthetic spectra to the observed ones in CH and CN spectral regions.
Abundances for C and N were determined together in an iteractive way, 
because for the temperature of our stars, carbon and nitrogen form molecules and as a 
consequence their abundances are related to each other.   
The input model atmosphere was used within the MOOG running {\em synth} driver that computes a set of trial synthetic spectra 
at higher resolution (0.3\AA~intervals) in the spectral region between 4150-4450\AA,
varying the carbon abundance in steps of 0.1  dex typically in the range
of --0.2 to --1.2 dex to fit a full band profile.
After the synthesis computations, the generated spectra were convolved with Gaussians of appropriate 
FWHM to match the resolution of the observed spectra.
In this way, the carbon abundances were derived by minimizing 
the observed-computed spectrum difference and were used 
to determine A(C). The carbon abundance was then used 
as input in the synthesis of the $UV$ CN feature to derive nitrogen abundances. 
The procedure was repeated until we obtained convergence within a tolerance of 0.1 dex in the C and N abundances.
\begin{figure}

\resizebox{\hsize}{!}{\includegraphics{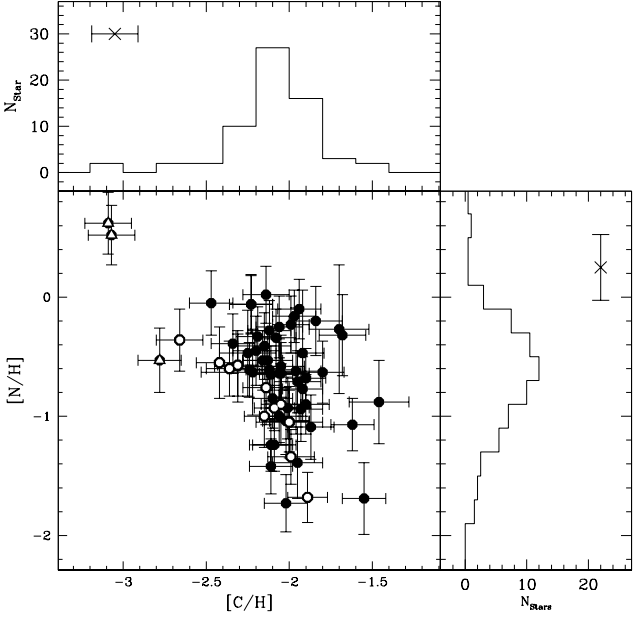}}
\caption{[C/H] and [N/H] abundances for the NGC1851 SGB, TO and MS stars in Table~\ref{stellar-par1} are plotted. 
A C versus N anticorrelation is evident. 
Stars that have already experienced some mixing episode are shown as large white dots, while three notable outliers are represented as
large triangles (see text for comments).
Histograms of [N/H] and [C/H]  with typical median error bars are also plotted in this figure.}
        \label{abbondanze}
   \end{figure}

For the results presented here, a fit was determined  by minimizing the observed-computed spectrum difference
in a 60\AA~window centered on 4300\AA~for the CH $G$-band and 40\AA~window for the $UV$ CN feature at 3883\AA.
Running {\em synth} on quite a broad spectral range  (200 and 300\AA~for the $G$-band and the CN feature, respectively)
to produce synthetic spectra allowed us to set a reasonable continuum level also by visual inspection and 
thus compute robust abundances.

We adopted a constant oxygen abundance ([O/Fe]=0.4 dex) throughout all computations.
The derived C abundance is dependent on the O abundance and therefore so is the N abundance, and in molecular 
equilibrium an over-estimate in oxygen produces an over-estimate of carbon (and vice versa), and an over-estimate 
of carbon from CN features is refleced in an under-estimate of nitrogen. 
We expect that the exact O values will affect the derived C abundances only negligibly,
since the CO coupling is marginal for stars warmer than $\sim$ 4500 K.
To test the sensitivity of the C abundance to the adopted O abundance we varied the oxygen abundances 
and repeated the spectrum synthesis to determine the exact dependence for a few representative stars 
in a wide range of $T_{eff}$ (from 5200 to 5900 K). 
In these computations, we adopted [O/Fe]= -0.5 dex,  [O/Fe]= 0.0 dex,  [O/Fe]= +0.5 dex.
We found that strong variations in the oxygen abundance slightly affect ($\delta A(C)/\delta [O/Fe] \simeq$ 0.15 dex) 
the derived C abundance in colder stars ($T_{eff}$$\leq$ 5400 K), while they are completely negligible 
(on the order of 0.05 dex or less) for warmer stars. This is within the uncertainty assigned to our measurement.

The total error in the A(C) and A(N) abundance was computed by taking into account the two main sources of uncertainty:
(i) the error in the adopted $T_{eff}$, typically $\delta$A(C)/$\delta T_{eff} \simeq$ 0.08--0.10 dex  and $\delta$A(N)/$\delta T_{eff} \simeq$ 0.11--0.13 per 100 K for the warmest stars in our sample\footnote{Cooler stars are slightly less sensitive to $T_{eff}$ variations, typically  at a level
of $\delta$A(C)/$\delta T_{eff} \simeq$ 0.07--0.08 dex and  $\delta$A(C)/$\delta T_{eff} \simeq$ 0.09-0.11 dex per 100 K.};
(ii) the error in the fitting procedure and  errors in the abundances that are 
likely caused by noise in the spectra\footnote{Additionally, in the treatment of internal error for nitrogen 
we varied the carbon abundance by $\pm$0.10 dex (that is the typical error associated to A(C)). We added these errors in quadrature 
with those introduced by the model atmosphere to estimate the internal uncertainty of the A(N) values.}.
The errors due to uncertainties on gravity and microturbulent velocity are negligible (on the order of 0.02 dex or less).
The sensitivity of the derived abundances to the adopted atmospheric parameters was obtained by repeating our abundance analysis
and changing only one parameter at each iteration for several stars that are representative of the  
temperature and gravity range explored. Thus, we assigned the internal error to each star depending on its  $T_{eff}$  and $\log g$.
The errors derived from the fitting procedure  were then added in quadrature to the errors introduced by 
atmospheric parameters, resulting in 
an overall error of $\sim \pm$0.14 dex for the C abundances and  $\sim \pm$0.28 dex for the N values.

Very recently  \citet{carretta10} found in NGC~1851 a small spread in metallicity  for a large number of giants, which is 
compatible with the presence of two different groups of stars whose metallicity differs by 0.06-0.08 dex 
\citep[but this result was not confirmed in][]{villanova10} .
This finding could affect our analysis in principle, because we adopted the same metallicity for all our stars 
in the synthesis ([Fe/H]=-1.22).
To test this effect we repeated the synthesis by altering the metallicity of stars belonging to the 
faintest SGB by 0.10 dex \citep[well above the spread claimed by][]{carretta10}.
The resulting abundance variations are within the uncertainty assigned to our measurement (typically $\delta A(C)/\delta [Fe/H] \simeq$ 0.07 dex  and $\delta A(N)/\delta [Fe/H] \simeq$ 0.04 dex) for our low-resolution 
spectra. Therefore this potential small [Fe/H]  variation among our spectroscopic targets has no influence on our analysis and conclusions.
We present the abundances derived as described above and the relative uncertainties in the abundance 
determination in Tab.~\ref{stellar-par1}. Additionally, this table lists the derived atmospheric parameters of all our targets.

\subsection {C and N abundance results}\label{abbond}

Fig.~\ref{abbondanze} plots [C/H] versus [N/H] measured in this paper.
We observe strong star-to-star variations in both elements, as already observed in all GCs studied to date.
An anticorrelation, with considerable scatter, is apparent from Fig.~\ref{abbondanze}.
The scatter is consistent with the observational errors, but there are a few outliers.
In a sample of 64 objects with Gaussian errors, two outliers at the 3 $\sigma$ level are not expected.
The deviation of stars 41350 (V = 19.3) and 40022 (V=19.9) with extremely depleted C, from the mean relation shown by the NGC~1851
sample in Fig.\ref{abbondanze} is of higher statistical significance.
We cannot provide a reliable explanation for this. Both stars are from the \citet{pancino10} sample and, 
judging from their radial velocities, are cluster members. Moreover, their $V,I$ magnitudes do not have large errors. As a tentative hypothesis we suggest that these two stars could belong to the {\em extreme}
population, using the scheme suggested by \citet{carretta09}.

\begin{figure}
\resizebox{\hsize}{!}{\includegraphics{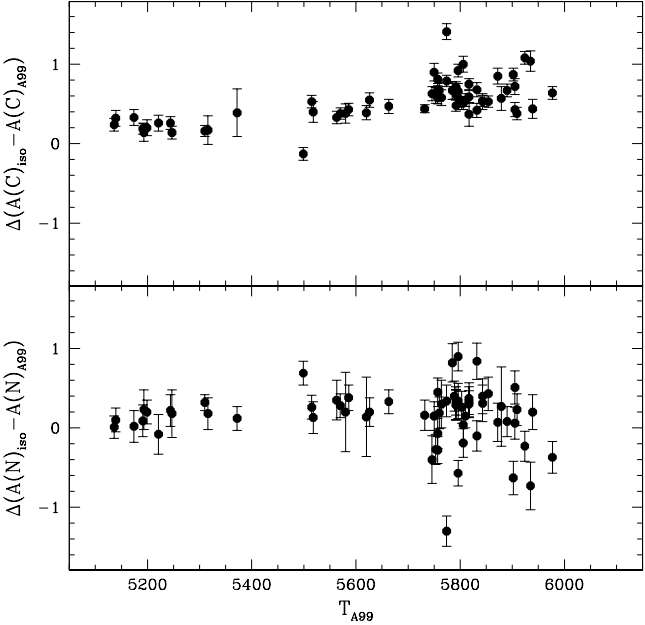}}
\caption{Comparison of  carbon (top) and nitrogen (bottom panel) abundances derived by adopting different temperature scales ("A99" refers to the \citealp{alonso99} calibration and "iso" to the isochrones fitting procedures) with their relative uncertainties.}
        \label{gratto}
   \end{figure}

All our stars are C-depleted, with moderately weak variations in carbon abundances (from [C/H]$\sim$--2.7 to [C/H]$\sim$--1.5) 
anticorrelated with strong variations in N. 
The nitrogen abundance spans almost 2 dex, from [N/H]$\sim$--1.9 up to [N/H]$\sim$0.0 dex\footnote{if we exclude the outliers discussed above.}.
To check the dependence of the carbon and nitrogen abundances on the adopted temperature scale, we re-ran the synthesis using
the atmospheric parameters derived by isochrone fitting (see Sect.~\ref{parametri}). The result of this exercise is shown in Fig.~\ref{gratto}. As can be seen from this figure, the carbon abundances would be higher considering
these higher temperatures (ranging from $\sim$ 0.2  dex for giants up to 0.4-0.5 dex for Ms stars). This reflects on the nitrogen abundances, as demonstrated in the bottom panel of the same figure. This is only to show that the abundances ranking among target stars is left unchanged: while the zero point of our derived abundances would shift, the amplitude of the star-to-star variations for C and N would remain similar regardless of the adopted stellar parameter. Therefore, our conclusions do not depend on the adopted stellar parameters.

\begin{figure}
\resizebox{\hsize}{!}{\includegraphics{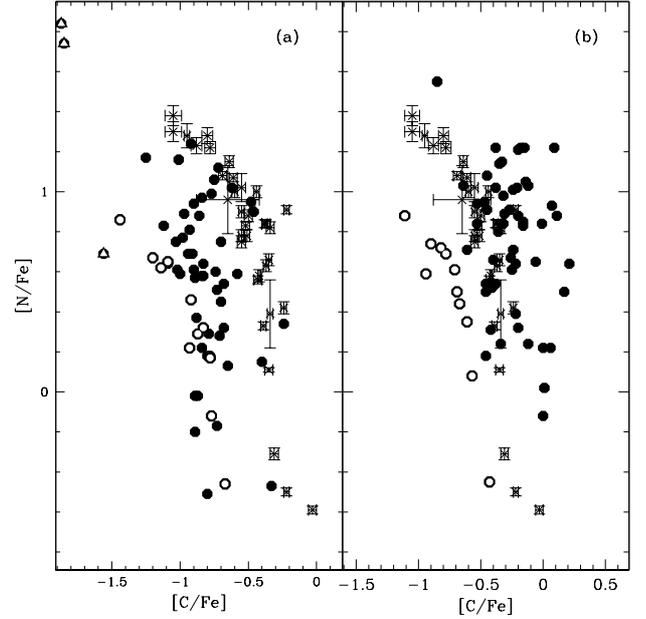}}
\caption{{\em Panel (a)}: [C/Fe] and [N/Fe] abundances for  NGC~1851 stars. Stellar atmospheres and spectral syntheses were derived by assuming the  \citet{alonso99} temperature scale. The symbols are the same as in Fig.~\ref{abbondanze}. 
Abundances and relative uncertainties for  stars in M~5 from  \citet{cohen02} are also shown as crosses for comparison. {\em Panel (b)}:  the same as in the left panel, but assuming temperatures and gravities obtained from the isochrones.}
        \label{cohen}
   \end{figure}

Evaluating the accuracy of our absolute abundance scale is very difficult because we found no literature 
data to compare with.  Fig.~\ref{cohen} compares the C and N abundances of this paper to the abundances derived for 
M~5 by \citet{cohen02}, a cluster with a metallicity comparable \citep[$\sim$--1.29;][2010 edition]{harris96} to that of NGC~1851.
In panel {\em (a)} we present a comparison between C and N abundances derived by assuming the \citet{alonso99} 
temperature scale and abundances derived by \citet{cohen02} for stars at the base of the RGB in M~5, while panel {\em (b)} refers to 
isochrones-fitting temperatures. In both panel a clear C-N anticorrelation is apparent.
According to theoretical computations and earlier investigations,  the carbon abundance declines from MS to RGB.
In panel {\em (a)} there is a mild disagreement with the  \citet{cohen02} data;
that is completely reconciled in panel {\em (b)}. 
This effect can be entirely explained because \citet{cohen02} used atmosphere parameters obtained from the isochrone. Still, although there is an offset, the two anticorrelations seem to follow a similar pattern. We conclude again from Fig.~\ref{cohen} that the anticorrelation we observe is totally untouched by the choice of the temperature scale, and shifts in the absolute abundance scale cannot account for the wide range in N abundances apparent in Fig.\ref{abbondanze}.

We therefore conclude that the C versus N anticorrelation among unevolved NGC~1851 stars in Fig.\ref{abbondanze}
is indeed real and from here on we will therefore only present results based on the \citet{alonso99} temperature scale.

\begin{figure}
\centering
\resizebox{\hsize}{!}{\includegraphics{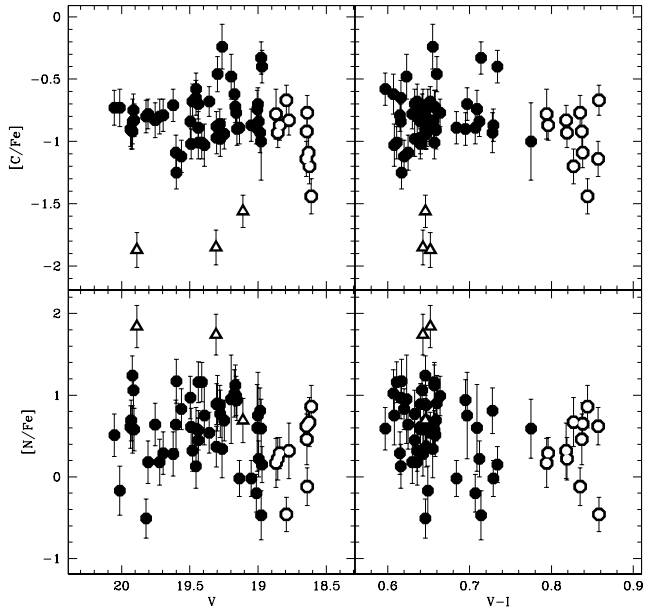}}
\caption{Derived C and N abundances are plotted against the photometry for NGC~1851 stars.
No systematic trends with either luminosity or temperature are apparent in the abundances.
We call stars marked as black empty triangles {\em anomalous}, while stars that 
supposedly underwent some mixing episodes are plotted as empty dots (see text).}
        \label{trend}
   \end{figure}

We also plotted the derived abundances as a function of the $V$ magnitude and $V-I$
color in Fig.~\ref{trend} to evaluate possible systematic effects with luminosity and temperature.

While none of these effects are apparent, we can tentatively identify the occurrence of 
a mixing episode for NGC~1851 stars from this plot. The top panel of Fig.~\ref{trend} shows a notable decline in the carbon
abundances for stars with $V\lesssim$18.9 and $(V-I)\gtrsim$0.8 (stars marked as white dots in the same figure),
which is expected for stars in the course of normal stellar evolution.
This behavior of the C abundance allows us to identify stars that experienced
a major mixing episode, which may alter the {\em primordial} abundances.
Curiously enough, these stars, plotted again as large white dots, seem to define a pretty clear and narrow 
anticorrelation in Fig.\ref{abbondanze} 
(Spearman's rank correlation coefficient --0.92). The shape of this anticorrelation agrees with 
what we expect after the occurrence of a mixing episode: the 
high N enhancement found in unevolved or less-evolved stars is strongly softened by evolutionary effects 
and a large part of dwarfs and early subgiants have N abundances 
as high as those observed in slightly evolved RGB stars.
We identify, again from Fig.~\ref{trend}, three outliers, coded as empty triangles as in Fig.~\ref{abbondanze}. 
Two of these stars were found to
deviate significantly from the main C-N relation if Fig.~\ref{abbondanze}. 
We call these {\em anomalous} only by virtue of their positions in the upper panel of Fig.~\ref{trend} and 
decided to not consider them further.

At this point we note that we cannot arbitrarily distinguish between two groups of stars with different 
[N/H] or [C/H] because we are unable to detect any clear bimodality.
To be more quantitative, we ran the dip test on unimodality \citep{1985}.
We performed this simple statistical test only on stars with a magnitude $V<$18.9 and
can confirm that there is no bimodality in either the [C/H] or [N/H].
\begin{figure}

\resizebox{\hsize}{!}{\includegraphics{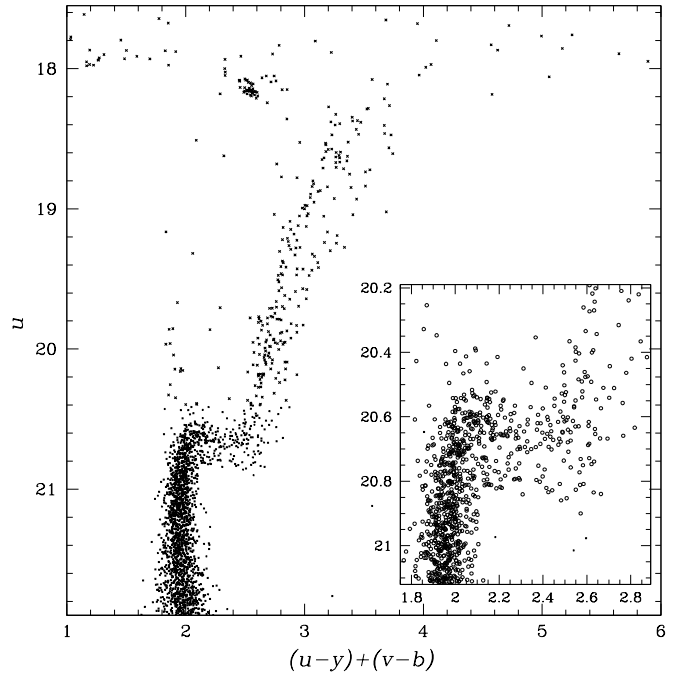}}
\caption {CMD for NGC~1851 from Str\"omgren photometry. 
The inset shows a zoom of the SGB region.
Only stars with high-quality photometry were plotted (see Sect.~\ref{fotometria} for details). 
Note the discrete double RGBs {\em connected} with the bimodal SGBs.}
 \label{CMD}
\end{figure}

\section{The chemical composition of the double RGB and SGB}
As already discussed in Sect.~\ref{introduzione}, the discovery of multiple sequences in the CMD of NGC~1851 provided  
unambiguous proof of the presence of multiple populations and brought new interest and excitement 
about this GC. 
While it is now widely accepted that NGC 1851 hosts two or more stellar populations, the connection among
its multiple SGBs, RGBs, and HBs is still controversial and the chemical composition of the two SGBs is also
debated.

\begin{figure}

\resizebox{\hsize}{!}{\includegraphics{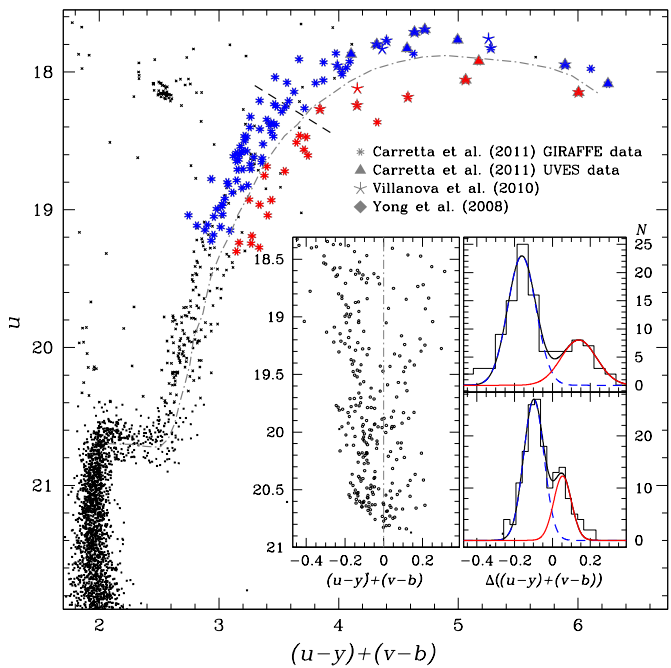}}
\caption {{\em Main panel}: Reproduction of the CMD of Fig.~\ref{CMD}. We used red color codes 
for red-RGB and faint-SGB stars, while blue-RGB and bright-SGB stars are
represented in blue. Symbols refer to stars from spectroscopic studies as
indicated in the figure. The rectified CMD and the histogram color distribution
of RGB stars between the two dashed lines is shown in the inset 
(see text for details).}
 \label{rapporto}
\end{figure}

Several authors suggested that the groups of $s$-rich and $s$-poor stars detected from RGB-star spectroscopy 
are the progeny of the faint- and bright-SGB, respectively \citep[e.\ g.\ M08,][]{yong08}, in close analogy 
to what was observed in M22 and $\omega$ Centauri \citep[e.\ g.][]{marino09,marino11,johnson10,pancino11}.
In contrast, \citet{carretta11} claimed that the faint-SGB consists of barium-poor metal-poor stars 
while bright-SGB stars have an enhanced barium and iron abundance. 

\subsection{Photometric connection between SGB and RGB}\label{fotometria}
To investigate this question in more detail, we started analyzing literature photometry. 
We used the WFC/ACS {\it HST} CMD in F606W and F814W bands presented in M08
\citep[see][for details]{sarajedini07,anderson08} and the 
 Str\"omgren $u$, $b$, $v$, $y$ photometry from \citet{grundahl99} and \citet{calamida07}.  
Here we are interested in high-quality photometry and included in the analysis only relatively isolated,
 unsaturated stars with good values of the PSF-quality fits and small rms errors in astrometry and photometry.
A detailed description of the selection procedures is given in \citet[Sect.~2.1]{milone09}. 
We corrected our photometry for remaining spatially dependent errors, caused by small
inaccuracies of the PSF model \citep[see][]{anderson08}.
To account for the color differences of these variations we followed the recipes from \citet[Sect.~3]{milone11a}.
Briefly, we  defined  a fiducial
line for the MS by computing a spline through the median colors found in
successive short intervals of magnitude, and we iterated this step with a
sigma clipping; then we examined the color residuals relative to the fiducial
and estimate for each star, how the observed stars in its vicinity may 
systematically lie  to the red or  the blue of  the fiducial sequence.
Finally we corrected the star's color by the difference between its color residuals.

To study multiple populations from the CMD analysis, we started searching for the combination of 
magnitude and colors that provides the best separation of the two RGBs and SGBs in NGC 1851. 
Results are illustrated in Fig.~\ref{CMD} where we plot $u$ as a function of ($u-y$)$+$($v-b$). 
A visual inspection of this diagram leaves no doubts on the presence of a bimodal RGB and SGB and shows that 
the faint-SGB and the bright-SGB are clearly connected with the red- and the blue-RGB, respectively.
A similar connection between the two SGBs and RGBs has already been 
observed  for NGC 1851  by \citet{han09} in the $U$ versus ($U-I$) CMD and was studied more recently
by \citet{sbordone11}. These authors showed that while the double SGB
is consistent with two groups of stars with either an age difference of about one Gyr or 
with different C+N+O overall abundance, the double RGB seems to rule out the possibility
of a large age difference.

Fig.~\ref{CMD} revealed that the bimodality found in the SGB can also be seen in the RGB.
To further confirm our association of the faint-SGB (bright-SGB) component with the red-RGB (blue-RGB),
 we focused on the relative number of stars of all evolutionary stages.
To estimate the fraction of stars in the two RGBs we used only RGB stars between the two dashed lines in the magnitude interval 
where the split is more evident (main panel of Fig.~\ref{rapporto}). The procedure is illustrated in the inset of Fig.~\ref{rapporto}.
 To obtain the straightened RGB of the right-hand panel, we subtracted from the color of each 
star the color of the fiducial sequence at the $u$ magnitude of the star. The color distribution 
of the points in the middle panel were analyzed in two magnitude bins.
The distributions have two clear peaks, which
we fitted with two Gaussians (red for the red-RGB and blue for the blue-RGB).
From the areas below the Gaussians, 70$\pm$3\% of stars turn out to belong to the blue-RGB, and 
30$\pm$3\% to the red one.
With the statistical uncertainties these fractions are the same in both magnitude intervals and roughly match 
the relative frequency on the fainter/brighter SGBs (35\% versus\ 65\%, M08) and HB stars on the 
blue/red  side of the instability strip (35\% versus \ 65\%).  

\setcounter{table}{2}
\begin{table*}
\caption{\label{tabella}Mean abundances for NGC~1851 stars from high-resolution studies.} 
\centering 
\begin{tabular}{l c c c c c} 
\hline\hline 
Element & Abundance (blue-RGB) & $N_{\rm stars}$ & Abundance (red-RGB) & $N_{\rm stars}$ &  References \\       
\hline 
$[$La/Fe$]$ 	&  0.27$\pm$0.02 	&  5 	 & 0.61$\pm$0.05  	& 3  &	1 \\
$[$Na/Fe$]$ 	&  --0.05$\pm$0.11	&  5  & 0.57$\pm$0.15	& 3 	& 	1 \\ 
$[$O/Fe$]$ 	&  0.50$\pm$0.04  	&  5  & 0.17$\pm$0.14  	& 3  & 	1 \\
$[$Fe/H$]$ 	& --1.29$\pm$0.04   &  5  & --1.23$\pm$0.07    & 3  & 	1  \\
$[$Ba/Fe$]$ &  0.09$\pm$0.02 &  8 &   0.52$\pm$0.03  & 7  &  2 \\
$[$Na/Fe$]$ &  0.04$\pm$0.11 &  8 &  0.47$\pm$0.07  & 7  &  2  \\
$[$O/Fe$]$ &  0.09$\pm$0.07 &  8  & --0.19$\pm$0.08  & 7  & 2 \\
$[$Fe/H$]$ & --1.23$\pm$0.01 &  8  & --1.22$\pm$0.01  & 7  & 2  \\
$[$Ba/Fe$]$ &  0.43$\pm$0.02 &  72 & 0.78$\pm$0.04  & 21 & 3\tablefootmark{a} \\
$[$Na/Fe$]$ &  0.13$\pm$0.03 &  81 &  0.47$\pm$0.04  & 24 & 3\tablefootmark{a} \\
$[$O/Fe$]$ &  0.04$\pm$0.02 &  66  & --0.14$\pm$0.05  & 17 & 3\tablefootmark{a}  \\
$[$Fe/H$]$ & --1.16$\pm$0.01 &  82  & --1.15$\pm$0.01  & 24 & 3\tablefootmark{a} \\
$[$Ba/Fe$]$ &  0.51$\pm$0.04 &  8 &  0.94$\pm$0.09  & 3  & 3\tablefootmark{b}  \\
$[$Na/Fe$]$ &  0.20$\pm$0.07 &  8  & 0.57$\pm$0.11  & 3  & 3\tablefootmark{b} \\
$[$O/Fe$]$ &  0.12$\pm$0.08 &  8  & --0.27$\pm$0.13  & 3  & 3\tablefootmark{b} \\
$[$Fe/H$]$ & --1.18$\pm$0.01 &  8  & --1.14$\pm$0.08  & 3  & 3\tablefootmark{b} \\

\hline 
\end{tabular}
\tablebib{
(1)~\citet{yong08}; (2) \citet{villanova10}; (3) \citet{carretta11}.
}
\tablefoot{
Red- and blue-RGB stars are defined according to their location with respect to the ridge line used to define color residuals in Fig.~\ref{rapporto}.
\tablefoottext{a}{GIRAFFE data}
\tablefoottext{b}{UVES data}
}
\end{table*}

\subsection {Chemical composition of NGC~1851 subpopulations}
Because the chemical abundance determinations presented so far do not define any clear bimodality,
the clear separation of the sequences of Fig.~\ref{CMD} 
provides a unique opportunity to obtain information on the 
chemical differences between the two RGBs and SGBs in NGC 1851. 
To do this, we used a  $u, (u-y)+(v-b)$ diagram to isolate
the samples of blue-RGB and bright-SGB stars, and red-RGB and faint-SGB stars.
Then we plotted with red and blue symbols the red-RGB and blue-RGB stars for which abundance measurements are 
available from high-resolution spectroscopy.

\begin{figure}
\resizebox{\hsize}{!}{\includegraphics{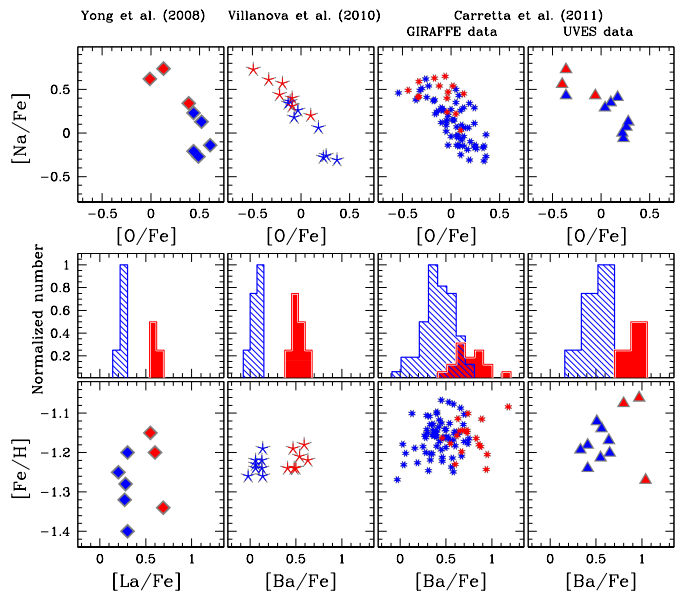}}
\caption {{\em Upper panel}: Na-O anticorrelation among NGC~1851 RGB stars from HR spectroscopy studies. 
Red color refers to stars photometrically 
selected to belong to the red-RGB in the Str\"omgren $u$,   $(u-y)+(v-b)$ diagram, while stars 
located on the blue-RGB are shown in blue. Symbols and color code are consistent with those of Fig.~\ref{rapporto}.
{\em Bottom panel}: The run of [La/Fe] versus [Fe/H] and the normalized number distribution for red and blue stars 
in this plane. The color code is consistent with the upper panel.}
 \label{letteratura}
\end{figure}

\begin{figure}
\centering
\resizebox{\hsize}{!}{\includegraphics{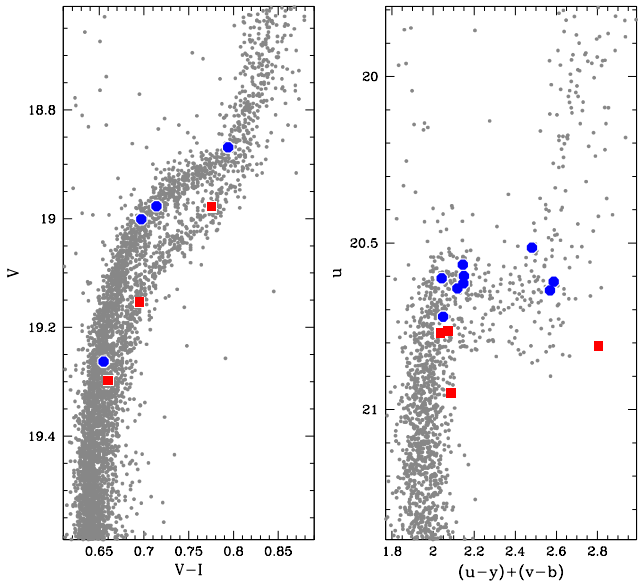}}
\caption{{\em Left panel:} selected bright-SGB (marked as blue dots) and faint-SGB (shown as red squares) 
stars are overplotted on the $V,V-I$ CMD presented by M08.
{\em Right panel:} selected bright- and faint-SGB stars are shown in the Str\"omgren $u$,  $(u-y)+(v-b)$ diagram.
The color code is consistent with the left panel. }
        \label{acs}
\end{figure}

Our analysis of the chemical abundance patterns of the two RGBs is summarized in Fig.~\ref{letteratura}.
Lower panels show [Fe/H] versus the abundances of the $s$-process elements barium and lanthanum measured by 
\citet{yong08}, \citet{villanova10}, and \citet{carretta11} 
from GIRAFFE and UVES data.
The histogram of the $s$-element distribution is illustrated in the middle panel, while upper panels plot
[Na/Fe] versus [O/Fe].
The average iron, barium, lanthanum, sodium, and oxygen abundances are listed in Table~\ref{tabella}
for the two groups of stars.

\begin{figure}
\centering
\resizebox{\hsize}{!}{\includegraphics{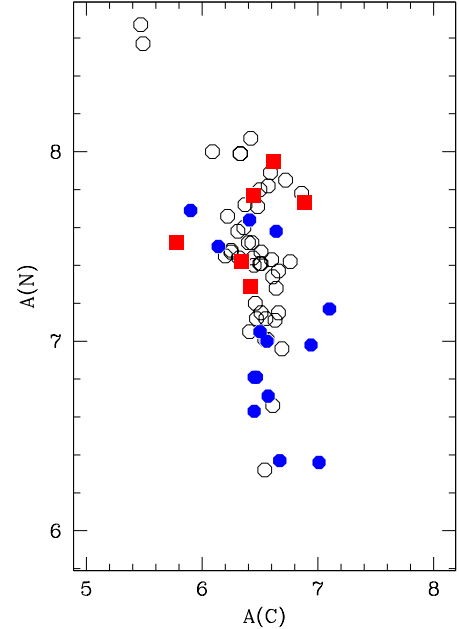}}
\caption{Photometrically selected bright-SGB (marked as blue dots) and faint-SGB (shown as red squares) stars are represented 
in the A(C) versus A(N) plane. Stars that we were unable to unambiguously associate with either of the two populations are shown with open circles.}
\label{pattern}
\end{figure}

In the light of our analysis of literature photometric and spectroscopic data we are now able to characterize
the two RGBs and SGBs of NGC 1851 as follows: 
\begin{itemize}
\item Faint-SGB and red-RGB  stars are photometrically connected, therefore they represent the same subpopulation of NGC~1851; 
the same can be said about bright-SGB and blue-RGB \citep[see also][for the case of M~22]{marino12}.
This connection is supported by the relative (percentage) 
numbers of the sequences; therefore the data do not support the interpretation by \citet{carretta10} that the red-RGB 
is associated to the bright-SGB.
\item Literature data  suggest that the red-RGB stars tend to be enriched on average in Na and $s$-process elements, 
and poor in oxygen, while blue-RGB stars appear to have their own, extended anticorrelation and to be solar 
in Ba and $s$-process elements. This is particularly evident in the \citet{carretta11} dataset, 
which also has the highest statistical value.
\item Red-RGB (and thus faint-SGB, according to our interpretation above) stars are enhanced in barium and 
lanthanum by $\sim$0.3-0.4 dex with respect to the blue-RGB 
(and consequently the bright-SGB). 
\item The literature data suggest that there is no significant iron difference between the two groups of stars. 
In this context we recall that \citet{yong08} and \citet{carretta11} detected a 
significant [Fe/H] variation among both $s$-rich and $s$-poor stars but these results 
strictly disagree with the narrow iron distribution observed by Villanova and collaborators. 
The presence of an intrinsic iron spread among NGC 1851 stars is still controversial.     
\end{itemize}

\subsection{C and N abundances along the double SGB}\label{cmds}

In this section we present the chemical composition of stars on the two SGBs of NGC 1851. 
A bona fide sample of stars that belong unambiguously either to the faint-SGB or to the bright-SGB were selected 
using both the $V$, $V-I$ and $u$,  $(u-y)+(v-b)$ diagrams (see Fig.~\ref{acs}).

The C and N abundances of these bona fide stars are plotted in 
Fig.~\ref{pattern}, where it is immediately clear that stars belonging to the bright-SGB show a fully developed 
anticorrelation, while stars belonging to the faint-SGB appear to have a smaller scatter, and to have on average an 
excess of N  (this remains still valid when considering  temperatures derived by isochrone fitting as input of the synthesis, as anticipated in Sect.~\ref{abbond}).
This new result supports our previous identification of the faint-SGB as the parent population of the red-RGB, and 
of the bright-SGB as the parent of the blue-RGB, not only on the basis of photometry and population ratios, but also 
on the basis of chemical composition.

\begin{figure}
\centering
\resizebox{\hsize}{!}{\includegraphics{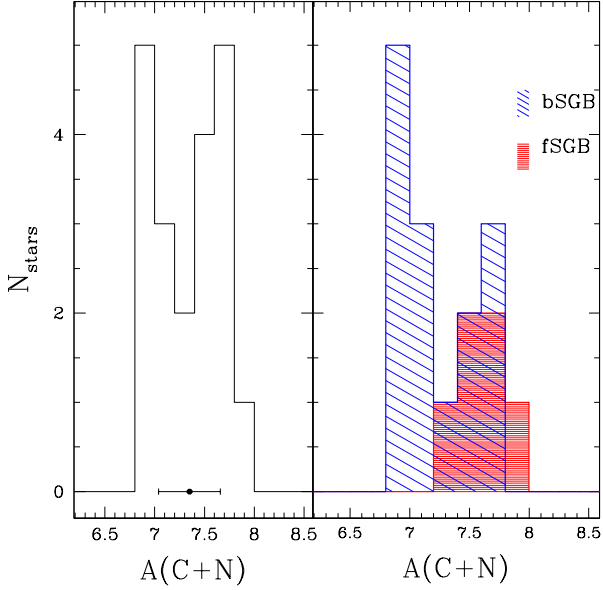}}
\caption{{\em Left panel:} Histogram of the C+N distribution for SGB stars selected as described in Sect.~\ref{cmds}. 
The median error bar is plotted below the histogram for reference.
 {\em Right panel:} Histograms of the C+N sum for bright-SGB (dashed blue) stars and faint-SGB (hatched red) stars are shown.}
        \label{istogrammi}
   \end{figure}

\citet{cassisi08} and \citet{ventura09} argued that the overall CNO abundance difference can account for the SGB split.
\citet{yong09} found evidence for strong CNO variations in contradiction with the the results of \citet{villanova10}. To further investigate this hypothesis,
we computed the C+N sum for our bona fide faint-SGB and bright-SGB stars.
We then derived the histograms of the distribution of the C+N sum.
In the left panel of Fig.~\ref{istogrammi}, we plot the histograms for the entire dataset (faint-SGB + bright-SGB bona fide stars) with the typical (median) error bar indicated.
The histogram shows a high dispersion with a hint of bimodality (with two clumps separated at A(C+N)$\simeq$7.35). For this larger dataset, according to a KMM test \citep{ashman94}, a bimodal distribution is a statistically significant improvement over the single Gaussian at a confidence level of 89\%.
However, a much clearer result is obtained when histograms are built considering the bright- and the faint-SGBs 
{\em separately} (right panel of Fig.~\ref{istogrammi}): they have a different (averaged) C+N content, the faint-SGB
having A(C+N)$\simeq$7.64$\pm$0.24 and the bright-SGB A(C+N)$\simeq$7.23$\pm$0.3 (A(C+N)$\simeq$7.80$\pm$0.19 and 
A(C+N)$\simeq$7.47$\pm$0.26; using isochrone fitting temperatures, respectively, see Sect.~\ref{abbond}). 
As an additional check, 
we performed a two-sample KS test computing the probability that these two samples are drawn from the same parent 
distribution and found a pretty high ($P_{KS}$ $\sim$0.03) significance of the difference in the bright-SGB and 
faint-SGB distribution of A(C+N).
From Fig.~\ref{pattern} bright-SGB stars appear to be on average more N-poor than then faint-SGB stars; assuming 
an N-O anticorrelation, the  bright-SGB stars  are more O-rich than the faint-SGB stars.
Even though we do not provide oxygen abundances for our SGB stars, we can speculate on the C+N+O sum for the two SGB components
assuming [O/Fe] values from available measurements.
Various oxygen abundance determinations of RGB stars can be found in the literature.
From Fig.~\ref{letteratura} we just note that large systematic differences exist between different determinations of the 
O content (see also Tab.~\ref{tabella}) and we caution readers that assigning a reference [O/Fe] 
content to each SGB group could be na\"ive at this stage.
If we assume for the bright and faint component [O/Fe]$\simeq$0.1 dex and [O/Fe]$\simeq$--0.2 dex, 
respectively\footnote{we derived these averaged values from the works of \citet{villanova10} and \citet{carretta10} 
reported in Table~\ref{tabella}.}, the separation one sees in C+N content virtually disappears when 
considering C+N+O. We found that the faint-SGB have A(C+N)$\simeq$7.89$\pm$0.14 and the bright-SGB A(C+N)$\simeq$7.93$\pm$0.07. 
The distributions  even {\em swap} when considering O 
abundances suggested by \citet{yong08} ([O/Fe]$\simeq$0.5 dex and [O/Fe]$\simeq$0.2 dex for the bright- and faint-SGB stars, 
respectively\footnote{here we note that our N abundances are systematically lower than \citet{yong09}, in some case as much as 
0.6-0.7 dex. We can attribute this discrepancy to (a) the different spectral resolution, (b) the different evolutionary 
status of program stars \citep[see Fig.~10 of][]{gratton00} and (c) the fact that in \citet{yong09} 
N measurements come from the CN features at 8005~\AA.}. We conclude that the bimodality we observe in the C+N sum 
does not necessarily imply or exclude a bimodality in the C+N+O content and more observations are needed to settle the 
case of NGC~1851.

\section{Summary and discussion}\label{conclusioni}

We presented low-resolution spectroscopy for a large sample of MS and SGB stars in  NGC~1851 with the goal
of deriving C abundances  (from the G band of CH) and N abundances (from the CN band at $\sim$ 3883\AA) and 
investigating the chemical differences between the two branches of the double SGB.
We derived carbon and nitrogen abundances for 64 stars, whose spectra were obtained with FORS2 at 
VLT and IMACS at Magellan and analyzed in a uniform manner.

NGC~1851 is one of the most interesting GCs 
whose CMD displays a discrete structure at the level of the SGB and of the RGB. 
The photometric complexity is reflected in a peculiar chemical pattern that has only recently been investigated in detail.
So far, the available abundance studies in NGC~1851 were limited to evolved stars that  
belong to the RGB \citep[except for the study of][]{pancino10}.
This is the first time that a precise chemical tagging of C and N content 
is made for stars directly located in the bright- and faint-SGB component. 
The main results of our analysis can be summarized as follows:

\begin{itemize}
\item We derived CH and CN band index measurements for 23 stars observed with IMACS, the spectrograph on the 
Magellan I telescope. We added to our sample spectra from \citet{pancino10}.
We were able to detect a large scatter and a hint of bimodality in the CN band strengths toward the brighter luminosities
(refer to Fig.~\ref{ridge}). We did not report any clear anti-corellation from these index measurements (Fig.~\ref{isto}).
\item We performed spectral synthesis  to separate the underlying C and N abundances from the CH and CN band strengths. 
Star-to-star strong variations with a significant range in A(C) and especially in A(N) were found at all luminosities from 
the MS  ($V\sim20.1$) up to the lower RGB ($V\sim18.6$).
C and N abundances are strongly anticorrelated, as would be expected from the presence of CN-cycle processing 
exposed material on the stellar surface (Fig.~\ref{abbondanze}). 
\item We used literature photometry in $u$, $b$, $v$, and $y$ Str\"omgren  bands \citep{grundahl99,calamida07} 
to define a new color index ($(u-y)+(v-b)$). We found that the $u$ versus   $(u-y)+(v-b)$ diagram 
is a powerful tool to identify the double RGB and SGB of NGC~1851 and showed that the faint-SGB
is clearly connected with the red-RGB, while blue-RGB stars are linked to the 
bright-SGB (Fig.~\ref{CMD}, see also \citealp{han09}). Moreover, the relative frequency on the 
fainter/brighter SGBs (35\% versus 65\%) roughly matches the relative frequency of red-/blue-RGB stars 
selected in the $u,(u-y)+(v-b)$ diagram (30\% versus 70\%; see Fig.~\ref{rapporto}). 
\item We {\em photometrically} defined blue- and red-RGB stars according to their position on this bimodal RGB sequence. 
We used $s$-elements, Na, O, and iron abundance that are available from literature for some 
RGB stars of both populations to investigate their chemical content. 
The less populous red-RGB consists of Ba-rich La-rich stars and have, on average, 
a higher Na abundance, while the bulk of Ba-poor La-poor stars belong to the blue-RGB.
However, since we have demonstrated that the two RGB and SGB are {\em photometrically} connected, we can confidently 
extend these results to the two SGB components for these $s$-process elements not studied in this paper. 
\item Similarly, we isolated bona fide stars on the faint-SGB and bright-SGB using 
available photometry and analyzed their chemical composition. We noted a fully extended C-N anticorrelation 
for the bright-SGB stars, while faint-SGB stars tend to be richer in N, on average (Fig.~\ref{pattern}). 
The C-N pattern observed for SGB stars recalls the Na-O anticorrelation analyzed for RGB stars 
in previous papers. Specifically, the faint-SGB and bright-SGB samples are not completely 
superimposed on one another in the A(C), A(N) plane; but faint-SGB stars have, on average, a higher nitrogen abundance. 
This finding rules out the claims by \citet{carretta11}, who suggested that the faint-SGB is also nitrogen-poor.
\item We analyzed the C+N sum for both bright-SGB and faint-SGB bona fide stars. Bright-SGB stars 
have A(C+N)$\simeq$7.23$\pm$0.31 dex, while  for the faint component 
A(C+N)$\simeq$7.64.$\pm$0.24 dex. A difference in $ \log_{\epsilon}$(C+N) of 0.4 dex as we 
find implies that the fainter SGB has about 2.5 times the C+N content of the brighter one. 
According to the \citet{cassisi08} scenario, the faint-SGB is anticipated to have the higher 
CNO content. The current findings of increased C+N content in the faint-SGB relative to 
the brighter one agree, in part, with the \citet{cassisi08} results.
However, we caution that 
the separation one sees in C+N content could significantly decrease or
disappear when considering the C+N+O sum (as discussed in Sect.~\ref{cmds}).
\end {itemize}

The general picture demonstrates that NGC~1851 shows an impressive resemblance to M~22.

M 22 possesses a spread in s-process elements, iron content (although this is still debated for NGC 1851), 
and each of the two populations exhibits its own anticorrelation, with the $s$-rich having on average 
higher C, N, and Na abundances. 
The chemical anomalies point to a bimodal SGB and RGB both for M22 and NGC 1851. 
Similarly to NGC~1851, also for M22 the faint-SGB and the bright-SGB consist
of $s$-rich and $s$-poor stars \citep[see][]{marino12}.

Since the Na-O and the C-N anticorrelations  {\em alone} can be considered as the signature 
of multiple stellar populations,
and both clusters are composed of two groups of stars with different $s$-element 
content (associated to the double SGB and RGB) possibly with their own Na-O, C-N 
anticorrelations, we conclude that each group in turn is the product of multiple 
stellar formation episodes.

NGC~1851 and M~22 do not harbor only two stellar populations (like {\em normal} GC) but 
have experienced a much more troubled star-formation history that resembles the 
case of $\omega$Centauri \citep[see e.\ g.\ discussions in][]{marino09,dacosta09,dacosta11,roederer11,dantona11},

\citet{dantona11} suggested for $\omega$ Centauri 
a chemical evolutionary scenario where due to the large mass of the proto-cluster and its possible dark matter halo 
the material ejected by SNII
may survive in a torus that collapses back onto the cluster after the SN II
epoch \citep[see also][]{dercole08}.
The 3D-hydro simulations by \citet{marcolini06} show indeed that the collapse back includes the matter enriched by the SN II ejecta.
This scenario could be easily extended to M22 and NGC~1851 \citep[see][]{marino_OMEGA}. 
For $\omega$ Centauri and M 22 it is tempting to speculate that enrichments in N and Na and depletion 
of C and O may have originated from the ejecta, collected in a cooling flow, of AGB stars that 
evolve in the cluster when the gas has been entirely exhausted by previous star-formation events. 

\citet{dantona11} suggested that a poorly discussed site of
$s$--nucleosynthesis that occurs in the carbon burning shells of
the tail of lower mass progenitors of SNII \citep[e.g.][]{the07},
may become particularly apparent in the evolution of the
progenitor systems of $\omega$~Cen, and similarly M22 and NGC~1851
\citep[see also][]{roederer11}.
 
As an alternative possibility, NGC~1851 has been recently suggested to be the merger-product of 
two independent stellar aggregates \citep{vanden96}. While this possibility seems unlikely 
for globular clusters in the Galactic halo, an origin as a merger product of two independent 
star clusters cannot be excluded in dwarf galaxies. In this case, numerical simulations \citep{bekki11} 
showed that two clusters can merge and form the nuclear star cluster of a dwarf galaxy. 
After the parent dwarf galaxy is accreted by the Milky Way, its dark matter halo and stellar envelope 
can be stripped by the Galactic tidal field, leaving behind the nucleus (i.e., NGC~1851) and a 
diffuse stellar halo \citep[as observed by][]{ol09}.

As already mentioned in the introduction, \citet{carretta11} associated 
the  $s$-rich and the $s$-poor  populations to the bright-SGB and the faint-SGB, 
respectively, with the bright-SGB having also higher N and Na abundances. 
According to Carretta and collaborators, the possibility that the faint-SGB is CNO 
enhanced should be excluded, demonstrating
that the split is caused by an age difference of $\sim$ 1 Gyr between the two populations. 
In this paper we have shown instead that the faint-SGB is made of N-rich and probably $s$-rich stars
and bright-SGB stars are N-poor and probably $s$-poor. While we added important pieces 
of information to the general picture, our results do not provide a conclusive answer on 
the occurrence of a merger in NGC 1851 and suggest that the measurement
of the overall C+N+O abundance as well as a precise determination 
of the spatial distribution of the multiple SGBs and RGBs are still 
mandatory to shed light on the star-formation history of this GC.

\begin{acknowledgements} 
We thank the anonymous referee for helpful comments, which greatly improved and clarified this work.
APM and RC acknowledges the funds by the Spanish Ministry of Science and Innovation under 
the Plan Nacional de Investigaraci\'on cient\'{\i}fica, Desarrollo e Investigaci\'on Tecnol\'ogica, AYA2010-16717. 
MZ acknowledges funding from the FONDAP Center for Astrophysics 15010003,
the BASAL CATA PFB-06, the Milky Way Millennium Nucleus from the
Ministry of Economycs ICM grant P07-021-F, and Proyecto FONDECYT Regular 1110393.
CL thanks the Istituto de Astrof\`\i sica de Canarias for its hospitality while parts of this paper were being completed.
This publication makes use of data products from the Two Micron All Sky Survey, which is a joint project of the University of Massachusetts and the Infrared Processing and Analysis Center/California Institute of Technology, funded by the National Aeronautics and Space Administration and the National Science Foundation. This research has made use of the SIMBAD database, operated at CDS, Strasbourg, France and of NASA�s Astrophysical Data System.

\end{acknowledgements} 

\bibliographystyle{aa}
\bibliography{bibliography.bib}

\longtab{1}{
\begin{longtable}{l c c c c c c c c c c }
\caption{\label{indici_tab}Index measurements for NGC~1851 stars} \\            
\hline \hline
     &                 &             &        &         &         &               &                  &          &             & \\ 
 ID  &      Ra	       &   Dec	     &     V  &  (V-I)  &    CN   &    $err_{CN}$ &     $\delta$ CN  &CH	&  $err_{CH}$ &   $ \delta$ CH   \\ 
     &  (deg)  	       & (deg)       &        &     	&         &   (mag) 	  &  (mag)           &  (mag)   &  (mag)      &  (mag) \\   
\hline
 &                 &             &        &         &         &               &                  &          &             & \\ 
\endfirsthead
\caption{continued.}\\
\hline\hline
     &                 &             &        &         &         &               &                  &          &             & \\ 
 ID  &      Ra	       &   Dec	     &     V  &  (V-I)  &    CN   &    $err_{CN}$ &     $\delta$ CN  &CH	&  $err_{CH}$ &   $ \delta$ CH   \\ 
     &  (deg)  	       & (deg)         &        &     	&         &   (mag) 	  &  (mag)           &  (mag)   &  (mag)      &  (mag) \\   
\hline
 &                 &             &        &         &         &               &                  &          &             & \\ 
\endhead
\endfoot
    
  11219  &  78.5334822 & -40.0274459 & 18.643 &  0.837 & $\cdots$ &    $\cdots$ &   $\cdots$ &     $\cdots$  &   $\cdots$ &   $\cdots$     \\
  11755  &  78.5401486 & -40.0328686 & 18.774 &  0.818 &  -0.381  &    0.047    &  -0.006    &      0.808    &     0.046  &   0.038   	   \\
  12485  &  78.5406833 & -40.0253661 & 19.001 &  0.697 &  -0.237  &    0.058    &   0.134    &      0.738    &     0.054  &  -0.038   	   \\
  12925  &  78.5447857 & -40.0234854 & 19.297 &  0.660 &  -0.261  &    0.063    &   0.106    &      0.734    &     0.062  &  -0.042   	   \\
  13062  &  78.5512037 & -40.0301395 & 18.624 &  0.827 &  -0.380  &    0.056    &  -0.002    &      0.781    &     0.039  &   0.017   	   \\
  13872  &  78.5527462 & -40.0445307 & 19.263 &  0.655 &  -0.371  &    0.063    &  -0.004    &      0.816    &     0.055  &   0.039   	   \\
  15182  &  78.5603478 & -40.0469688 & 18.646 &  0.857 & $\cdots$ &    $\cdots$ &   $\cdots$ &     $\cdots$  &   $\cdots$ &   $\cdots$	   \\
  15490  &  78.5599836 & -40.0488694 & 18.977 &  0.714 &  -0.388  &    0.038    &  -0.016    &      0.878    &     0.043  &   0.102   	   \\
  16047  &  78.5459885 & -40.0581902 & 18.610 &  0.844 &  -0.314  &    0.046    &   0.064    &      0.737    &     0.029  &  -0.026   	   \\
  20295  &  78.5276335 & -40.0632061 & 18.630 &  0.838 &  -0.207  &    0.031    &   0.171    &      0.810    &     0.040  &   0.046   	   \\
  40017  &  78.4595330 & -40.1525297 & 19.695 &  0.615 &  -0.406  &    0.023    &   0.020    &      0.679    &     0.043  &  -0.033  	   \\
  40020  &  78.4724650 & -40.1511010 & 18.841 &  0.796 &  -0.256  &    0.024    &   0.106    &      0.841    &     0.045  &   0.036  	   \\
  40022  &  78.4332081 & -40.1504760 & 19.888 &  0.652 &  -0.336  &    0.023    &   0.105    &      0.669    &     0.042  &  -0.053  	   \\
  40028  &  78.4695406 & -40.1494668 & 19.494 &  0.633 &  -0.418  &    0.023    &  -0.007    &      0.716    &     0.036  &   0.002  	   \\
  40051  &  78.4631667 & -40.1439444 & 18.792 &  0.796 &  -0.357  &    0.052    &   0.018    &      0.811    &     0.043  &   0.040   	   \\
  40062  &  78.4834487 & -40.1414047 & 19.161 &  0.664 &  -0.324  &    0.023    &   0.062    &      0.772    &     0.036  &   0.028  	   \\
  40072  &  78.4780543 & -40.1393568 & 19.934 &  0.648 &  -0.412  &    0.025    &   0.032    &      0.684    &     0.049  &  -0.042  	   \\
  40078  &  78.4568778 & -40.1381667 & 19.464 &  0.629 &  -0.413  &    0.051    &  -0.049    &      0.725    &     0.056  &  -0.048   	   \\
  40083  &  78.4307625 & -40.1371584 & 19.305 &  0.647 &  -0.390  &    0.033    &   0.007    &      0.674    &     0.045  &  -0.053  	   \\
  40088  &  78.4695640 & -40.1359246 & 19.248 &  0.647 &  -0.395  &    0.024    &  -0.003    &      0.680    &     0.042  &  -0.053  	   \\
  40094  &  78.4774387 & -40.1347787 & 19.490 &  0.640 &  -0.464  &    0.032    &  -0.053    &      0.693    &     0.045  &  -0.021  	   \\
  40097  &  78.4733785 & -40.1341169 & 20.053 &  0.658 &  -0.434  &    0.025    &   0.019    &      0.755    &     0.050  &   0.015  	   \\
  40100  &  78.4495611 & -40.1340000 & 18.924 &  0.751 &  -0.390  &    0.061    &  -0.017    &      0.799    &     0.039  &   0.024   	   \\
  40117  &  78.4580365 & -40.1314412 & 19.358 &  0.652 &  -0.345  &    0.026    &   0.056    &      0.720    &     0.049  &  -0.002  	   \\
  40123  &  78.4664750 & -40.1306111 & 19.006 &  0.709 &  -0.387  &    0.054    &  -0.016    &      0.716    &     0.042  &  -0.060   	   \\
  40133  &  78.4144056 & -40.1287500 & 19.193 &  0.667 & $\cdots$ &    $\cdots$ &   $\cdots$ &     $\cdots$  &   $\cdots$ &   $\cdots$	   \\
  40153  &  78.4534972 & -40.1264167 & 19.479 &  0.632 & $\cdots$ &    $\cdots$ &   $\cdots$ &     $\cdots$  &   $\cdots$ &   $\cdots$     \\
  40167  &  78.4457694 & -40.1245556 & 18.932 &  0.706 & $\cdots$ &    $\cdots$ &   $\cdots$ &     $\cdots$  &   $\cdots$ &   $\cdots$     \\
  40186  &  78.4047861 & -40.1218056 & 18.869 &  0.817 & $\cdots$ &    $\cdots$ &   $\cdots$ &     $\cdots$  &   $\cdots$ &   $\cdots$     \\
  40191  &  78.4702263 & -40.1212996 & 19.598 &  0.617 &  -0.380  &     0.036   &     0.039  &      0.699    &     0.048  &  -0.012  	   \\
  40196  &  78.4923929 & -40.1204995 & 19.649 &  0.622 &  -0.436  &     0.022   &    -0.013  &      0.662    &     0.040  &  -0.049  	   \\
  40197  &  78.4345480 & -40.1204489 & 19.412 &  0.657 &  -0.355  &     0.020   &     0.050  &      0.733    &     0.031  &   0.015  	   \\
  40235  &  78.5103649 & -40.1167899 & 19.921 &  0.646 &  -0.272  &     0.021   &     0.171  &      0.723    &     0.031  &  -0.002  	   \\
  40239  &  78.4986784 & -40.1165524 & 19.562 &  0.620 &  -0.415  &     0.019   &     0.001  &      0.651    &     0.036  &  -0.061  	   \\
  40241  &  78.4430254 & -40.1165272 & 19.693 &  0.622 &  -0.408  &     0.023   &     0.018  &      0.675    &     0.040  &  -0.037  	   \\
  40247  &  78.4682250 & -40.1159722 & 18.705 &  0.837 &  -0.375  &     0.039   &     0.001  &      0.814    &     0.058  &   0.046   	   \\
  40271  &  78.5131639 & -40.1140000 & 18.866 &  0.844 &  -0.248  &     0.055   &     0.126  &      0.925    &     0.040  &   0.152   	   \\
  40303  &  78.4411306 & -40.1118056 & 19.033 &  0.686 & $\cdots$ &    $\cdots$ &   $\cdots$ &     $\cdots$  &   $\cdots$ &   $\cdots$     \\
  40340  &  78.4730104 & -40.1093919 & 19.913 &  0.642 &  -0.419  &     0.019   &     0.023  &      0.724    &     0.037  &  -0.000  	   \\
  40344  &  78.5028991 & -40.1090554 & 19.438 &  0.633 &  -0.444  &     0.027   &    -0.037  &      0.726    &     0.037  &   0.010  	   \\
  40348  &  78.4456867 & -40.1089487 & 19.820 &  0.646 &  -0.371  &     0.027   &     0.064  &      0.724    &     0.037  &   0.007  	   \\
  40376  &  78.4770704 & -40.1067791 & 19.930 &  0.659 &  -0.401  &     0.034   &     0.043  &      0.754    &     0.055  &   0.028  	   \\
  40378  &  78.5259583 & -40.1070000 & 18.740 &  0.825 & $\cdots$ &    $\cdots$ &   $\cdots$ &     $\cdots$  &   $\cdots$ &   $\cdots$     \\
  40385  &  78.5091532 & -40.1061629 & 19.806 &  0.630 &  -0.449  &     0.020   &    -0.015  &      0.735    &     0.044  &   0.019  	   \\
  40424  &  78.4685664 & -40.1034580 & 19.438 &  0.657 &  -0.309  &     0.027   &     0.098  &      0.740    &     0.037  &   0.024  	   \\
  40431  &  78.5162111 & -40.1031389 & 18.881 &  0.811 & $\cdots$ &    $\cdots$ &   $\cdots$ &     $\cdots$  &   $\cdots$ &   $\cdots$     \\
  40465  &  78.4746278 & -40.1013333 & 18.976 &  0.757 &  -0.476  &     0.101   &    -0.104  &      0.733    &     0.156  &  -0.043   	   \\
  40504  &  78.4685451 & -40.0992400 & 19.454 &  0.597 &  -0.405  &     0.020   &     0.003  &      0.724    &     0.044  &   0.008  	   \\
  40507  &  78.4706722 & -40.0993611 & 19.173 &  0.607 &  -0.396  &     0.036   &    -0.027  &      0.752    &     0.051  &  -0.025   	   \\
  40508  &  78.5094543 & -40.0990863 & 19.602 &  0.625 &  -0.350  &     0.025   &     0.069  &      0.689    &     0.038  &  -0.022  	   \\
  40545  &  78.4757778 & -40.0975278 & 18.985 &  0.728 &  -0.380  &     0.035   &    -0.008  &      0.765    &     0.043  &  -0.011   	   \\
  40571  &  78.5047352 & -40.0960942 & 19.437 &  0.611 &  -0.321  &     0.027   &     0.086  &      0.707    &     0.039  &  -0.010  	   \\
  40575  &  78.4698407 & -40.0960196 & 20.013 &  0.649 &  -0.498  &     0.025   &    -0.048  &      0.740    &     0.042  &   0.005  	   \\
  40620  &  78.4532694 & -40.0945000 & 18.640 &  0.835 &  -0.458  &     0.030   &    -0.081  &      0.870    &     0.042  &   0.105   	   \\
  40665  &  78.5128445 & -40.0926724 & 18.995 &  0.712 &  -0.443  &     0.029   &    -0.070  &      0.765    &     0.037  &  -0.007  	   \\
  40679  &  78.4762083 & -40.0924444 & 18.716 &  0.803 &  -0.250  &     0.038   &     0.126  &      0.821    &     0.039  &   0.053   	   \\
  40709  &  78.4309537 & -40.0912317 & 19.166 &  0.657 &  -0.494  &     0.020   &    -0.108  &      0.707    &     0.044  &  -0.036  	   \\
  40715  &  78.4849583 & -40.0910365 & 19.908 &  0.644 &  -0.388  &     0.025   &     0.054  &      0.700    &     0.038  &  -0.024  	   \\
  40756  &  78.4788241 & -40.0895444 & 19.280 &  0.647 &  -0.427  &     0.036   &    -0.032  &      0.730    &     0.040  &   0.001  	   \\
  40827  &  78.5001750 & -40.0876944 & 18.762 &  0.836 & $\cdots$ &    $\cdots$ &   $\cdots$ &     $\cdots$  &   $\cdots$ &   $\cdots$     \\
  40863  &  78.5070126 & -40.0863515 & 19.393 &  0.608 &  -0.396  &     0.018   &     0.007  &      0.737    &     0.040  &   0.018  	   \\
  40874  &  78.4722925 & -40.0860825 & 19.455 &  0.616 &  -0.341  &     0.027   &     0.067  &      0.721    &     0.039  &   0.005  	   \\
  40919  &  78.5234044 & -40.0849874 & 19.496 &  0.637 & $\cdots$ &    $\cdots$ &   $\cdots$ &     $\cdots$  &   $\cdots$ &   $\cdots$     \\
  40978  &  78.4762989 & -40.0829703 & 19.919 &  0.640 &  -0.424  &     0.023   &     0.019  &      0.691    &     0.047  &  -0.034  	   \\
  41003  &  78.4946844 & -40.0825071 & 19.050 &  0.729 &  -0.495  &     0.029   &    -0.118  &      0.767    &     0.037  &   0.005  	   \\
  41018  &  78.5251795 & -40.0823269 & 18.855 &  0.819 &  -0.365  &     0.032   &     0.009  &      0.816    &     0.041  &   0.043   	   \\
  41108  &  78.4583652 & -40.0800752 & 19.754 &  0.656 &  -0.423  &     0.025   &     0.007  &      0.722    &     0.038  &   0.008  	   \\
  41185  &  78.5009451 & -40.0785878 & 19.110 &  0.646 &  -0.418  &     0.024   &    -0.036  &      0.709    &     0.035  &  -0.043  	   \\
  41213  &  78.5164942 & -40.0782312 & 19.154 &  0.695 &  -0.371  &     0.024   &    -0.002  &      0.725    &     0.038  &  -0.052   	   \\
  41279  &  78.4260692 & -40.0762239 & 19.277 &  0.633 &  -0.391  &     0.026   &     0.003  &      0.683    &     0.039  &  -0.047  	   \\
  41325  &  78.4670381 & -40.0752654 & 19.480 &  0.636 &  -0.360  &     0.027   &     0.050  &      0.702    &     0.040  &  -0.012  	   \\
  41350  &  78.4739954 & -40.0746754 & 19.308 &  0.643 &  -0.509  &     0.018   &    -0.112  &      0.663    &     0.039  &  -0.063  	   \\
  41372  &  78.4922083 & -40.0743611 & 18.777 &  0.856 & $\cdots$ &    $\cdots$ &   $\cdots$ &     $\cdots$  &   $\cdots$ &   $\cdots$     \\
  41558  &  78.4702879 & -40.0710717 & 19.721 &  0.636 &  -0.388  &     0.023   &     0.040  &      0.680    &     0.047  &  -0.033  	   \\
  41610  &  78.4540666 & -40.0700619 & 19.137 &  0.684 &  -0.454  &     0.036   &    -0.070  &      0.743    &     0.040  &  -0.004  	   \\
  41694  &  78.5000889 & -40.0687222 & 18.736 &  0.847 & $\cdots$ &    $\cdots$ &   $\cdots$ &     $\cdots$  &   $\cdots$ &   $\cdots$     \\
  41807  &  78.4921679 & -40.0666366 & 18.792 &  0.858 &  -0.367  &     0.025   &    -0.009  &      0.926    &     0.058  &   0.109  	   \\
  41835  &  78.4320835 & -40.0662831 & 19.012 &  0.707 &  -0.345  &     0.018   &     0.029  &      0.700    &     0.040  &  -0.068  	   \\
  41884  &  78.5039348 & -40.0656429 & 18.978 &  0.775 & $\cdots$ &    $\cdots$ &   $\cdots$ &     $\cdots$  &   $\cdots$ &   $\cdots$     \\
  42073  &  78.4802671 & -40.0628473 & 19.304 &  0.640 &  -0.418  &     0.021   &    -0.022  &      0.737    &     0.038  &   0.010  	   \\
  42195  &  78.4415574 & -40.0610284 & 19.622 &  0.643 &  -0.394  &     0.024   &     0.026  &      0.723    &     0.035  &   0.012  	   \\
  42623  &  78.4582694 & -40.0551667 & 18.971 &  0.734 &  -0.413  &     0.031   &    -0.041  &      0.715    &     0.049  &  -0.060   	   \\
  42785  &  78.4376004 & -40.0528185 & 19.496 &  0.616 &  -0.418  &     0.035   &    -0.007  &      0.722    &     0.043  &   0.008  	   \\
  42865  &  78.4964339 & -40.0519295 & 18.869 &  0.794 & $\cdots$ &    $\cdots$ &   $\cdots$ &     $\cdots$  &   $\cdots$ &   $\cdots$     \\
  43014  &  78.4127639 & -40.0498611 & 19.197 &  0.623 &  -0.379  &     0.047   &    -0.011  &      0.722    &     0.056  &  -0.055  	   \\

 \hline
\end{longtable}
}

\longtab{2}{
\begin{longtable}{l c c c c c c}
\caption{\label{stellar-par1}Atmospheric parameters and carbon and nitrogen abundances for NGC~1851 stars}\\            
\hline \hline
     &                 &             &        &         &         &                \\ 
 ID  &      $T_{eff}$  &   $\log g$  &  A(C)  &  eA(C)  &   A(N)  &    eA(N)       \\ 
     &       (K)       &             &        &     	&         &     	   \\   
\hline
     &                 &             &        &         &         &                 \\
\endfirsthead
\caption{continued.}\\
\hline\hline
     &                 &             &        &         &         &                \\ 
 ID  &      $T_{eff}$  &   $\log g$  &  A(C)  &  eA(C)  &   A(N)  &    eA(N)       \\ 
     &       (K)       &             &        &     	&         &     	   \\   
\hline
     &                 &             &        &         &         &                 \\
\endhead
\endfoot

 11219 &  5193 $\pm$   90  &  3.4  &  6.42 &  0.14  &  7.29  & 0.31  \\
11755 &  5247 $\pm$   93  &  3.5  &  6.51 &  0.12  &  7.15  & 0.34  \\
12485 &  5620 $\pm$  116  &  3.8  &  6.64 &  0.13  &  7.58  & 0.53  \\
12925 &  5746 $\pm$  124  &  3.9  &  6.88 &  0.14  &  7.73  & 0.34  \\
13062 &  5221 $\pm$   92  &  3.4  &  6.14 &  0.14  &  7.50  & 0.30  \\
13872 &  5764 $\pm$  125  &  3.9  &  7.10 &  0.18  &  7.17  & 0.35  \\
15182 &  5139 $\pm$   87  &  3.4  &  6.20 &  0.14  &  7.45  & 0.23  \\
15490 &  5563 $\pm$  112  &  3.7  &  7.01 &  0.13  &  6.36  & 0.30  \\
16047 &  5174 $\pm$   90  &  3.4  &  5.90 &  0.14  &  7.69  & 0.26  \\
20295 &  5191 $\pm$   90  &  3.4  &  6.25 &  0.12  &  7.48  & 0.26  \\
40017 &  5909 $\pm$  135  &  4.1  &  6.55 &  0.13  &  7.12  & 0.26  \\
40020 &  5310 $\pm$   97  &  3.6  &  6.47 &  0.12  &  7.12  & 0.19  \\
40022 &  5774 $\pm$  126  &  4.2  &  5.47 &  0.14  &  8.67  & 0.26  \\
40062 &  5732 $\pm$  123  &  3.9  &  6.57 &  0.11  &  7.82  & 0.24  \\
40072 &  5789 $\pm$  128  &  4.2  &  6.44 &  0.14  &  7.44  & 0.26  \\
40083 &  5792 $\pm$  127  &  3.9  &  6.46 &  0.12  &  7.20  & 0.25  \\
40088 &  5792 $\pm$  127  &  3.9  &  6.40 &  0.12  &  7.52  & 0.25  \\
40094 &  5817 $\pm$  129  &  4.0  &  6.32 &  0.12  &  7.44  & 0.23  \\
40097 &  5753 $\pm$  125  &  4.2  &  6.61 &  0.14  &  7.34  & 0.26  \\
40117 &  5774 $\pm$  126  &  4.0  &  6.66 &  0.12  &  7.37  & 0.25  \\
40123 &  5580 $\pm$  102  &  3.7  &  6.60 &  0.15  &  7.43  & 0.53  \\
40191 &  5902 $\pm$  135  &  4.1  &  6.09 &  0.13  &  8.00  & 0.27  \\
40197 &  5757 $\pm$  125  &  4.0  &  6.33 &  0.13  &  7.99  & 0.24  \\
40235 &  5796 $\pm$  128  &  4.2  &  6.42 &  0.14  &  8.07  & 0.24  \\
40239 &  5890 $\pm$  135  &  4.1  &  6.22 &  0.13  &  7.66  & 0.25  \\
40340 &  5810 $\pm$  129  &  4.2  &  6.59 &  0.13  &  7.89  & 0.22  \\
40344 &  5843 $\pm$  130  &  4.0  &  6.64 &  0.12  &  7.28  & 0.27  \\
40348 &  5796 $\pm$  128  &  4.2  &  6.54 &  0.13  &  6.32  & 0.24  \\
40376 &  5750 $\pm$  124  &  4.2  &  6.43 &  0.15  &  7.52  & 0.25  \\
40385 &  5854 $\pm$  132  &  4.2  &  6.56 &  0.12  &  7.01  & 0.26  \\
40424 &  5757 $\pm$  125  &  4.0  &  6.45 &  0.13  &  7.40  & 0.24  \\
40504 &  5977 $\pm$  140  &  4.1  &  6.76 &  0.13  &  7.42  & 0.26  \\
40507 &  5939 $\pm$  117  &  3.9  &  6.72 &  0.16  &  7.85  & 0.29  \\
40508 &  5872 $\pm$  133  &  4.1  &  6.25 &  0.14  &  7.47  & 0.30  \\
40545 &  5518 $\pm$  100  &  3.7  &  6.41 &  0.16  &  7.64  & 0.28  \\
40571 &  5924 $\pm$  137  &  4.0  &  6.33 &  0.13  &  7.99  & 0.25  \\
40575 &  5785 $\pm$  127  &  4.2  &  6.61 &  0.15  &  6.66  & 0.30  \\
40620 &  5199 $\pm$   86  &  3.4  &  6.57 &  0.14  &  6.71  & 0.23  \\
40665 &  5570 $\pm$  113  &  3.7  &  6.50 &  0.12  &  7.05  & 0.22  \\
40709 &  5757 $\pm$  125  &  3.9  &  6.62 &  0.13  &  7.95  & 0.25  \\
40715 &  5803 $\pm$  129  &  4.2  &  6.51 &  0.13  &  7.41  & 0.26  \\
40756 &  5792 $\pm$  127  &  3.9  &  6.48 &  0.12  &  7.71  & 0.24  \\
40863 &  5935 $\pm$  130  &  4.0  &  6.31 &  0.17  &  7.58  & 0.36  \\
40874 &  5905 $\pm$  135  &  4.0  &  6.69 &  0.14  &  6.96  & 0.27  \\
40978 &  5817 $\pm$  130  &  4.2  &  6.50 &  0.18  &  7.41  & 0.29  \\
41003 &  5515 $\pm$  109  &  3.7  &  6.47 &  0.13  &  6.81  & 0.22  \\
41018 &  5244 $\pm$   93  &  3.6  &  6.41 &  0.12  &  7.05  & 0.26  \\
41108 &  5760 $\pm$  125  &  4.1  &  6.51 &  0.14  &  7.47  & 0.26  \\
41185 &  5796 $\pm$  127  &  3.9  &  5.78 &  0.13  &  7.52  & 0.27  \\
41213 &  5626 $\pm$  116  &  3.8  &  6.44 &  0.13  &  7.77  & 0.25  \\
41279 &  5843 $\pm$  131  &  4.0  &  6.36 &  0.14  &  7.60  & 0.29  \\
41325 &  5832 $\pm$  130  &  4.0  &  6.66 &  0.14  &  7.15  & 0.25  \\
41350 &  5806 $\pm$  128  &  4.0  &  5.49 &  0.14  &  8.57  & 0.25  \\
41558 &  5832 $\pm$  130  &  4.1  &  6.54 &  0.14  &  7.01  & 0.28  \\
41610 &  5663 $\pm$  119  &  3.8  &  6.45 &  0.13  &  6.81  & 0.22  \\
41807 &  5136 $\pm$   87  &  3.5  &  6.67 &  0.12  &  6.37  & 0.21  \\
41835 &  5586 $\pm$  114  &  3.8  &  6.45 &  0.13  &  6.63  & 0.23  \\
41884 &  5372 $\pm$   97  &  3.7  &  6.34 &  0.31  &  7.42  & 0.36  \\
42073 &  5817 $\pm$  129  &  4.0  &  6.37 &  0.13  &  7.72  & 0.25  \\
42195 &  5806 $\pm$  128  &  4.1  &  6.63 &  0.13  &  7.11  & 0.27  \\
42623 &  5499 $\pm$   99  &  3.7  &  6.94 &  0.13  &  6.98  & 0.22  \\
42785 &  5905 $\pm$  130  &  4.1  &  6.50 &  0.14  &  7.80  & 0.26  \\
42865 &  5316 $\pm$   93  &  3.6  &  6.56 &  0.20  &  7.00  & 0.30  \\
43014 &  5879 $\pm$  115  &  3.9  &  6.86 &  0.18  &  7.78  & 0.54  \\
 \hline
\end{longtable}
}

\end{document}